\documentclass[acmsmall]{acmart}

\acmConference
\acmBooktitle

\AtBeginDocument{%
  }

\usepackage{xspace}
\usepackage{graphicx}
\usepackage{pifont}
\usepackage{multirow, makecell}
\usepackage[table,xcdraw]{xcolor}
\usepackage{subcaption}
\usepackage{bm}
\usepackage{enumitem}
\usepackage[ruled,vlined,linesnumbered]{algorithm2e}
\usepackage{environ} 

\usepackage{tcolorbox}
\tcbuselibrary{breakable}

\newcommand{\hi}[1]{\vspace{.25em} \noindent {\bf #1} }
\newcommand{\bfit}[1]{\textbf{\textit{#1}}}

\newcommand{\llm}{\textsc{LLM}\xspace}
\newcommand{\llms}{\textsc{LLMs}\xspace}
\newcommand{\oursys}{\bfit{DBCooker}\xspace} 

\newtheorem{definition}{DEFINITION}[section]
\newcommand{\gray}[1]{\textcolor{gray}{#1}}

\setcopyright{cc}
\setcctype{by}
\acmYear{2026}
\acmPrice{}
\acmDOI{10.1145/3802018}

\acmJournal{PACMMOD}
\acmVolume{4}
\acmNumber{3 (SIGMOD)}
\acmArticle{141}
\acmMonth{6}

\begin{document}

\title{Automating Database-Native Function Code Synthesis with LLMs}

\author{Wei Zhou}
\affiliation{\institution{{Shanghai Jiao Tong University}}\country{China}}
\email{{weizhoudb@sjtu.edu.cn}}

\author{Xuanhe Zhou}
\authornote{Xuanhe Zhou is the corresponding author.}
\affiliation{\institution{{Shanghai Jiao Tong University}}\country{China}}
\email{{zhouxuanhe@sjtu.edu.cn}}

\author{Qikang He}
\affiliation{\institution{{Shanghai Jiao Tong University}}\country{China}}
\email{{hqk0205@sjtu.edu.cn}}

\author{Guoliang Li}
\affiliation{\institution{{Tsinghua University}}\country{China}}
\email{{liguoliang@tsinghua.edu.cn}}

\author{Bingsheng He}
\affiliation{\institution{{National University of Singapore}}\country{Singapore}}
\email{{dcsheb@nus.edu.sg}}

\author{Quanqing Xu}
\affiliation{\institution{{Ant Group}}\country{China}}
\email{{xuquanqing.xqq@oceanbase.com}}

\author{Fan Wu}
\affiliation{\institution{{Shanghai Jiao Tong University}}\country{China}}
\email{{fwu@cs.sjtu.edu.cn}}

\renewcommand{\shortauthors}{Wei Zhou et al.}

\begin{abstract}
  Database systems incorporate an ever-growing number of functions built in their kernels (a.k.a., database native functions) for scenarios like new application support and business migration. This growth causes an urgent demand for automatic database native function synthesis. While recent advances in LLM-based code generation (e.g., Claude Code) show promise, existing approaches are too generic for database-specific development. They often hallucinate or overlook critical context because database function synthesis is inherently complex and error-prone, where synthesizing a single database function may involve registering multiple function units (e.g., for different input types), placing code in the correct source files, linking internal references, and implementing logic correctly.

  To this end, we propose \oursys, an \llm-based system for automatically synthesizing database native functions. The system consists of three key components. First, the \emph{function characterization module} aggregates multi-source declarations, identifies function units that require {specialized coding} through hierarchical analysis, and traces cross-unit dependencies via static analysis. Second, we design operations to address the main synthesis challenges: (1) a \emph{pseudo-code–based coding plan generator} that constructs structured implementation skeletons by identifying key elements such as reusable referenced functions; (2) a \emph{hybrid fill-in-the-blank model} guided by probabilistic priors and component awareness to integrate core logic with reusable routines; and (3) \emph{three-level progressive validation}, including syntax checking, standards compliance, and LLM-guided semantic verification. Finally, an \emph{adaptive orchestration strategy} unifies these operations with existing database tools and dynamically sequences them based on the orchestration history of similar functions. Results show that our system outperforms state-of-the-art methods on SQLite, PostgreSQL, and DuckDB ({34.55\%} higher accuracy on average), and can synthesize four categories of new functions absent in the latest SQLite (v3.50). 
  The code is available at \emph{{\url{https://github.com/weAIDB/DBCooker}}}.
\end{abstract}



\received{October 2025}
\received[revised]{January 2026}
\received[accepted]{February 2026}

\maketitle

\section{Introduction}
\label{sec:intro}







\begin{figure}[!t]
    \centering
    \vspace{1em}
    \begin{subfigure}{.85\textwidth}
    \hspace{.5em}
        \includegraphics*[width=\textwidth, trim={0em 1em 5em 0}, clip]{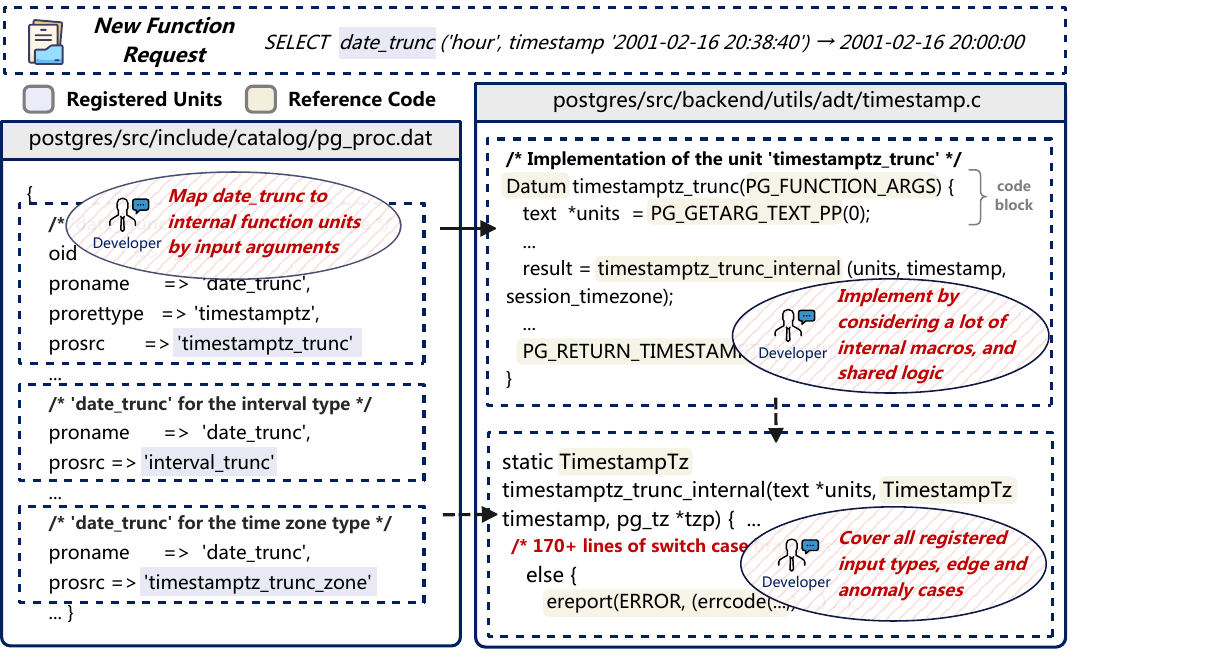}
        \caption{Example Synthesis by Human Developers}
    \end{subfigure}
    \begin{subfigure}{0.42\textwidth}
        \includegraphics*[width=\textwidth]{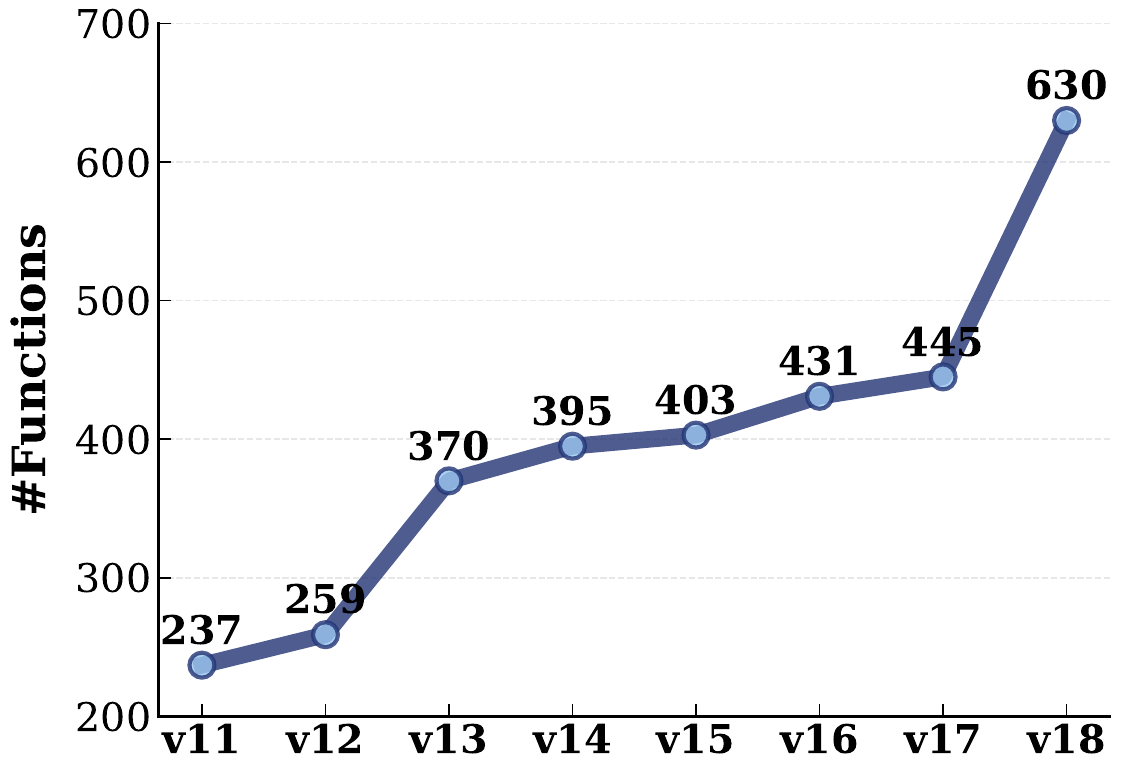}
        \caption{\#Functions over PG Version}
    \end{subfigure}
    \begin{subfigure}{0.46\textwidth}
        \includegraphics*[width=\textwidth]{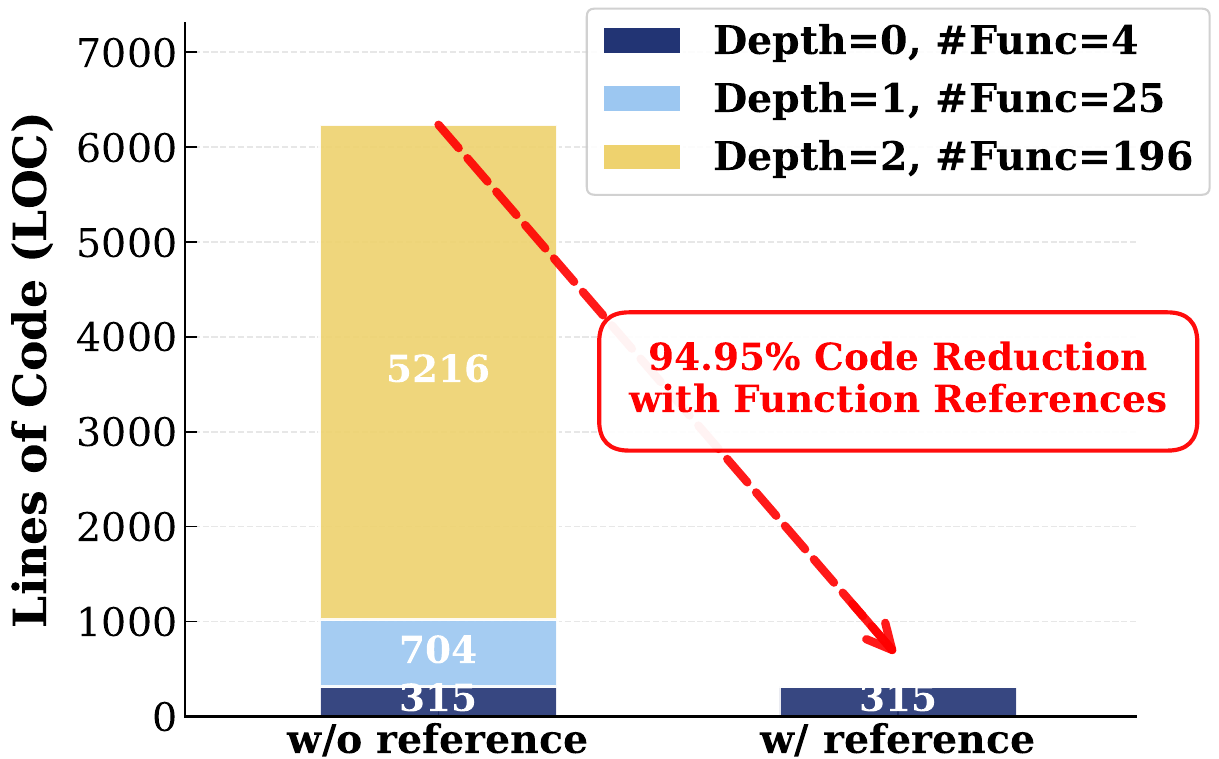}
        \caption{Code for Function References}
    \end{subfigure}
    \caption{{Database native function code synthesis is a complex problem} \normalfont{-- (a) Example synthesis workflow by human developers; (b) Increasing tendency of function number across versions; (c) Code size comparison with vs. without reference usage}.}
    \label{fig:intro}
    \vspace{-.5cm}
\end{figure}

Databases offer numerous {native SQL} functions in their kernels. For instance, PostgreSQL v18 includes 27 date functions such as \emph{date\_trunc()}, and DuckDB v1.4.0 offers 57 numeric functions such as \emph{sqrt()}.
The number of native SQL functions has been steadily increasing in modern database systems.
For example, as shown in Figure~\ref{fig:intro} (b), PostgreSQL functions nearly tripled from 237 (v11) to 630 (v18)~\cite{postgresqlfunction}, DuckDB grew from 60 (v0.3.3) to 666 (v1.4.0)~\cite{duckdbfunction}, and SQLite increased from 52 (v3.8.0) to 143 (v3.50.0)~\cite{sqlitefunction}.
This expansion is driven by new scenario support (e.g., BI analysis~\cite{abbasi2024adaptive, LLMDATASurvey, BID, zhoullmpreparationsurvey} and geometric processing~\cite{makris2019performance}) and business migration~\cite{migration, cracksql, DBAIOps}.
Specifically, in legacy migration scenarios (e.g., Oracle to PostgreSQL), implementing proprietary functions is a major bottleneck~\cite{migrationchallenges1, migrationchallenges2, OpenMLDBIndustry}, with code refactoring accounting for 30\%--60\% of migration budgets and requiring 40--80 hours per 1,000 code lines~\cite{migrationcost}.

Synthesizing database native functions is a critical task for extending system capabilities, as evidenced by the official development guides provided by major systems such as PostgreSQL~\cite{postgresqlextension, duckdbextension}.
However, it is a tightly coupled, error-prone process that demands extensive human expertise~\cite{douglas2003postgresql} and a deep understanding of internal dependencies and differences across major database updates~\cite{duckdbdependency, duckdbversion}, imposing a hard-to-sustain burden on developers.
The complexity is substantial: PostgreSQL native function codes span 119,161 lines (within two dependency hops) from v11 to v18, and the DuckDB GitHub repository reports 3,791 function-related issues~\cite{duckdb}.
As shown in Figure~\ref{fig:intro} (a), implementing the SQL function \emph{date\_trunc()} in PostgreSQL requires developers to (1) register appropriate function units (e.g., \emph{timestamptz\_trunc} and \emph{interval\_trunc}) based on factors such as {argument types} specified in the \emph{prorettype} attribute; (2) search for the correct source files to implement the function units (e.g., \emph{timestamp.c} for \emph{timestamptz\_trunc}); and (3) reference the correct units (e.g., \emph{PG\_GETARG\_TEXT\_PP}) to complete the code blocks in these function units. Failing to utilize these references, such as {misusing some internal data types}, can result in standard violations (causing synthesis failures) or re-implementation waste.
As shown in Figure~\ref{fig:intro} (c), implementing only the four registered units of \emph{date\_trunc()} using reference units reduces the required code lines by approximately 94.95\% compared to implementing it from scratch, which would require 6,235 lines of code and 225 functions across two-hop invocations.



\hi{Limitations of Existing Methods.}
Despite recent advances in \llm-based general-purpose code generation~\cite{llmcodesurvey}, to our knowledge, there is no publicly available tool or framework that provides a fully automated pipeline for database native functions.
Existing code synthesis methods, including prompt-based approaches~\cite{chen2021evaluating,copilot2023empirical}, agent-based systems~\cite{li2022alphacode,qian2023sweagent}, and training-based models~\cite{roziere2023codellama,li2023starcoder,luo2023wizardcoder}, demonstrate significant limitations in database native function synthesis.
First, these general-purpose methods fail to capture the specific characteristics of database native functions, such as required function units and fine-grained references within the repository, resulting in incorrect or incomplete implementations.
For example, Qwen Code~\cite{hui2024qwen2} may omit a function unit needed to handle interval inputs in \emph{date\_trunc()}.
Second, they rely heavily on coarse-grained file search operations without domain-specific heuristics to guide code placement, and overlook essential verification steps. 
For instance, Claude Code~\cite{claudecode} spends most of its time scanning unrelated files to locate where to implement \emph{date\_trunc()}, rather than validating its correctness across different input values.
Third, existing methods use a static synthesis strategy that starts with file exploration and implements functions from scratch, without accounting for differences in function complexity or leveraging similarities across functions. For example, they treat a simple math function \emph{sqrt()} and a complex aggregate function \emph{json\_agg()} identically, and fail to reuse related implementations such as \emph{date\_part()} of the same date function category  (see analysis in Section~\ref{subsec:pilot}).

\begin{sloppypar} 
\hi{Challenges.} It presents three main challenges. First, a high-level SQL function may correspond to multiple underlying function units with distinct names and responsibilities, making it non-trivial to determine which units need to be synthesized. For example, in PostgreSQL, {\emph{date\_trunc()} needs to register functions like \emph{timestamptz\_trunc} and \emph{interval\_trunc} separately for input arguments in timezone and interval}  (\textbf{C1}). 
Second, implementing a database native function requires referencing a vast number of existing function units, without which the implementation procedure cannot be correctly synthesized, because (1) the references are complex and cannot easily generate from scratch (e.g., the lines of code for \emph{data\_trunc()} references within two hops increase from 315 to 6,235 lines in total); and (2) some references are essential for functionality (e.g., output formatting using macros \emph{PG\_RETURN\_NUMERIC} in PostgreSQL) (\textbf{C2}).
Third, generalizing synthesis across diverse functions is difficult.
Simple functions such as \emph{sqrt()} can be implemented by directly wrapping standard libraries, whereas high-complexity aggregate functions, such as \emph{json\_agg()}, require custom logic, necessitating adaptive synthesis strategies (\textbf{C3}).
\end{sloppypar}

\hi{Our Methodology.} To address these challenges, we propose \oursys, an automatic LLM-based database function synthesis system. First, to accurately capture function composition and key information, we introduce the \textit{function characterization} module with (1) multi-source function declaration collection (e.g., documents and system catalogs), (2) {hierarchical function unit identification} to isolate distinctive units (require implementation), and (3) {context-aware cross-unit reference analysis} using static dependency graphs and category-specific pruning (\textbf{for C1}).
Second, to improve synthesis accuracy and correctly utilize references, we design \textit{synthesis operations}, including (1) {format-based pseudo-plan generation}, creating and scoring structured code skeletons to guide generation, (2) a {hybrid fill-in-the-blank coding model} enhanced with probabilistic priors and component awareness for precise incorporation of distinctive and reusable reference units, and (3) a \textit{three-level progressive validation} module, covering syntax, compliance, and \llm-guided checking (\textbf{for C2}).
Finally, to handle functions of varying complexity, we propose an  \textit{adaptive synthesis operation orchestration} module that abstracts operations as tools and dynamically optimizes the tool calling workflow using a combination of LLM-driven decision making and similar workflow trajectories (\textbf{for C3}).


\hi{Contributions.}  In summary, we make the following contributions.

\noindent $\bullet$ We propose an LLM-based system for automatic synthesis of database function code. {\it To our best knowledge, it is the first system that analyzes function composition, generates required internal references and linkages across functional units, and adaptively synthesizes various database functions} (Section~\ref{sec:overview}).


\noindent $\bullet$ We propose a function characterization module that captures key implementation information via declaration extraction, distinct unit identification, and cross-unit reference analysis (Section~\ref{sec:offline}). 

\noindent $\bullet$ We propose a distinctive function unit code synthesis strategy via format-based pseudo-plan generation, hybrid fill-in-the-blank model guided coding, and progressive code validation (Section~\ref{sec:synthesis}).

\noindent $\bullet$ We employ an adaptive diverse-function code synthesis mechanism that dynamically orchestrates synthesis operations via a hybrid optimization strategy combining \llm-based decision making and {trajectory-based} workflow reuse (Section~\ref{sec:orchestration}).

\noindent $\bullet$ We conduct extensive experiments to assess different synthesis methods over three mainstream databases and \oursys~outperforms state-of-the-art methods (e.g., Claude Code~\cite{claudecode}) with 34.55\% higher accuracy on average and can add completely new functions absent in the latest SQLite (v3.50) (Section~\ref{sec:exp}).
The code is available at \emph{{\url{https://github.com/weAIDB/DBCooker}}}.

\vspace{-.3cm}
\section{Preliminary}
\label{sec:preliminary}


In this section, we first introduce database native functions, and then formalize the problem of database function code synthesis. 


\vspace{-.25cm}
\subsection{Database Native Functions}
\label{subsec:databasefunction}

Database native functions provide a wide range of functionalities (e.g., string manipulation, numeric computation) and can be decomposed into three parts: (1) \emph{function declaration}, (2) \emph{new function units to be implemented}, and (3) \emph{existing function units used as references}.

\hi{Function Declaration.} Given a database engine $\mathcal{D}$, a \emph{function declaration} $\bm{\mathit{f_{dec}}}$ formally specifies the function's existence and usage, including the name, description, argument types, and return type.
The declaration is typically stored in the system catalog to support consistent registration and invocation.
For example, the declaration of \emph{date\_trunc()} describes its functionality ``truncate timestamp'', and specifies the two input arguments (\emph{text, timestamp}) in \emph{pg\_proc.dat}.

\hi{Function Unit.} A {function unit} $\bm{\mathit{f_{unit}}}$ is a self-contained {executable} component with one or multiple code blocks $\{\bm{\mathit{c_{block}}}\}$ {defined in specific database files}. Each code block (i.e., continuous code lines) encapsulates a specific aspect of the function's behavior, ranging from its main computational logic to auxiliary tasks (e.g., process the input arguments or format the output results). 

For function code synthesis, given an SQL function, we distinguish between the {newly synthesized function units} $\bm{\mathit{f^{new}_{unit}}}$ (which need to be implemented) and the {reference function units} $\bm{\mathit{f^{ref}_{unit}}}$ (which are already implemented, such as library functions or macros).

\begin{sloppypar}
For example, the \emph{date\_trunc()} function owns newly synthesized function units (e.g., \emph{timestamptz\_trunc}) defined in \emph{timestamp.c}, and multiple reference function units (e.g., \emph{PG\_GETARG\_TEXT\_PP}).
\end{sloppypar}




\hi{Database Native Function.} {Different from functions like UDFs} defined in SQL level, the {database native function} $\bm{\mathit{f}}$ is fully implemented by one or multiple function units and directly integrated into the database $\mathcal{D}$, which can be formally represented as below.

\vspace{-.5em}
\begin{sloppypar}
\begin{definition}[Database Native Function]\label{def:databasenativefunction}
A database native function $f$ defined in a database $\mathcal{D}$ is represented as $\langle \bm{\mathit{{f_{dec}}}}, \{ \langle \bm{\mathit{f^{new}_{unit}}}, \bm{\mathit{{f^{ref}_{unit}}}}\}\rangle \}  \rangle$, which is composed of one or multiple functions units $\{\bm{\mathit{f^{new}_{unit}}}\}$, where each $\bm{\mathit{f^{new}_{unit}}}$ can invoke a set of external {referenced} function units $\{\bm{\mathit{f^{ref}_{unit}}}\}$ (e.g., internal modules, macros), and is exposed through a unified SQL-level function interface defined in function declaration $\bm{\mathit{{f_{dec}}}}$. 
\end{definition}
\end{sloppypar}

\begin{sloppypar}
\begin{example}
Database native function \emph{date\_trunc()} in PostgreSQL includes: (1) $\bm{\mathit{f_{dec}}}$ that specifies the signature with two input arguments (\emph{text, timestamp}) and one output argument \emph{timestamp};
(2) $\bm{\mathit{f_{unit}}}$ with four newly synthesized function units (i.e., \emph{timestamp\_trunc}, \emph{interval\_trunc}, \emph{timestamptz\_trunc}, and \emph{timestamptz\_trunc\_zone}) defined in \emph{src/backend/utils/adt/timestamp.c} to handle different inputs, and multiple reference function units (e.g., \emph{PG\_GETARG\_TEXT\_PP} to get the input arguments, and \emph{PG\_RETURN\_TIMESTAMPTZ} to format output results).
\end{example}
\end{sloppypar}

\subsection{Database Native Function Code Synthesis}
\label{subsec:funcsynthesisdef}

We next define the problem of \emph{database native function code synthesis}. Given a precise \emph{function specification} (e.g., function name, description, input arguments, output types, and SQL examples), the goal is to {automatically generate and implement executable function codes within the database kernel} such that the synthesized code correctly realizes the intended functionality. 


\begin{sloppypar}
\begin{definition}[{Database Native Function Synthesis}] 
Given a {SQL-level} function specification $\mathcal{S}$ for the target database $\mathcal{D}$, database native function synthesis aims to generate the codes of {necessary function units} $\{\bm{f^{new}_{unit}}\}$ that satisfy $\mathcal{S}$ and can be successfully integrated into $\mathcal{D}$ with all the essential references $\{\bm{f^{ref}_{unit}}\}$, i.e., without any compliance error and passing all the test-cases $\mathcal{T}$ in database $\mathcal{D}$ with expected results. 

\end{definition}
\end{sloppypar}


\begin{sloppypar}

\begin{example}
{\it To synthesize the \emph{date\_trunc()} function, the input declaration includes (1) the code-style function declaration with the name and the input arguments, (2) the textual declaration with the usage descriptions, and SQL examples. Accordingly, we first identify the reference function units such as \emph{PG\_FUNCTION\_ARGS} for processing input arguments.
We then implement multiple new function units of key processing, such as \emph{timestamptz\_trunc()} and \emph{timestamp\_trunc()}, which handle input validation and output construction.
Once all the function units are correctly implemented, the function can be executed using the SQL ``\emph{SELECT date\_trunc('hour', timestamp '2001-02-16 20:38:40')}'', producing the result ``2001-02-16 20:00:00''.}
\end{example}


\end{sloppypar}

\subsection{Pilot Study}
\label{subsec:pilot}

We characterize database native function codes and investigate the limitations of existing methods.

\begin{figure}[!t]
    \centering
    \begin{subfigure}{0.44\textwidth}
        \includegraphics*[width=\textwidth]{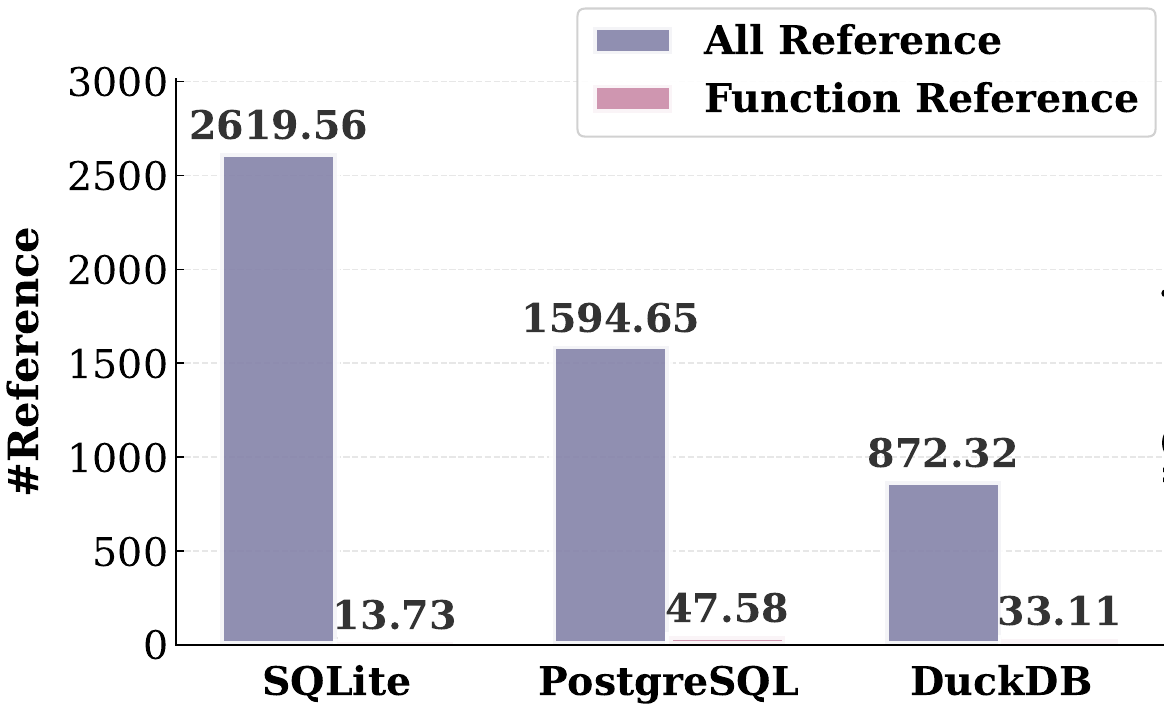}
        \caption{\#Reference within Database}
    \end{subfigure}
    \begin{subfigure}{0.44\textwidth}
        \includegraphics*[width=\textwidth]{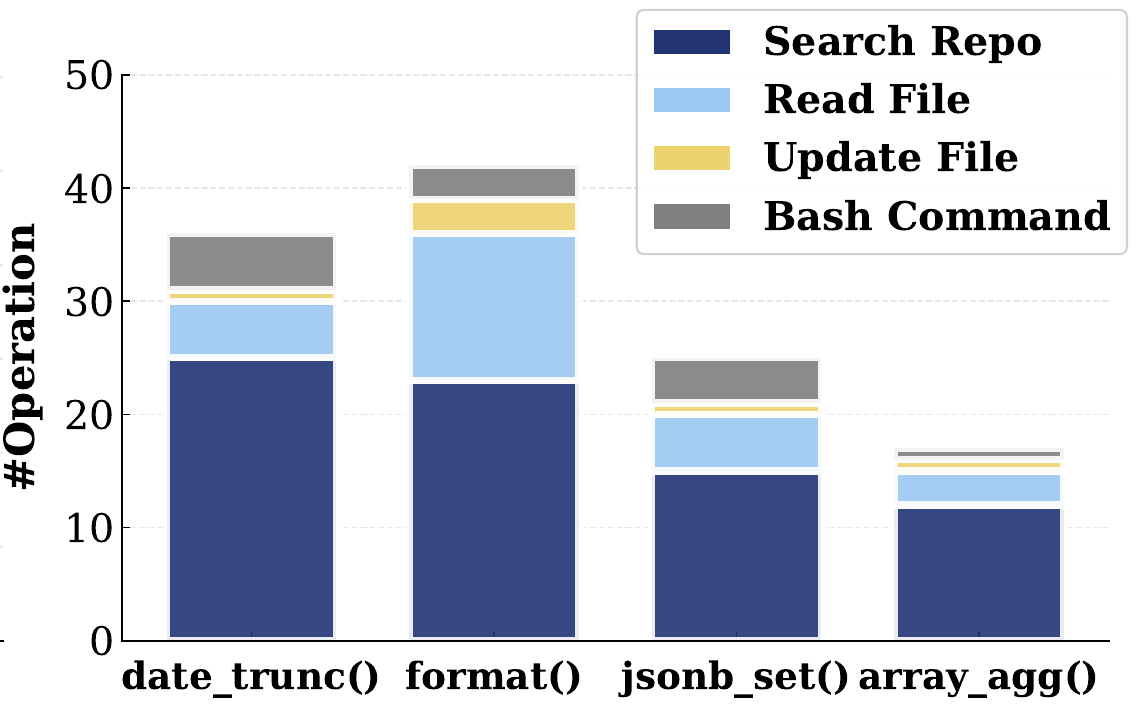}
        \caption{Operation Distribution}
    \end{subfigure}
    \begin{subfigure}{.85\textwidth}
        \includegraphics*[width=\textwidth]{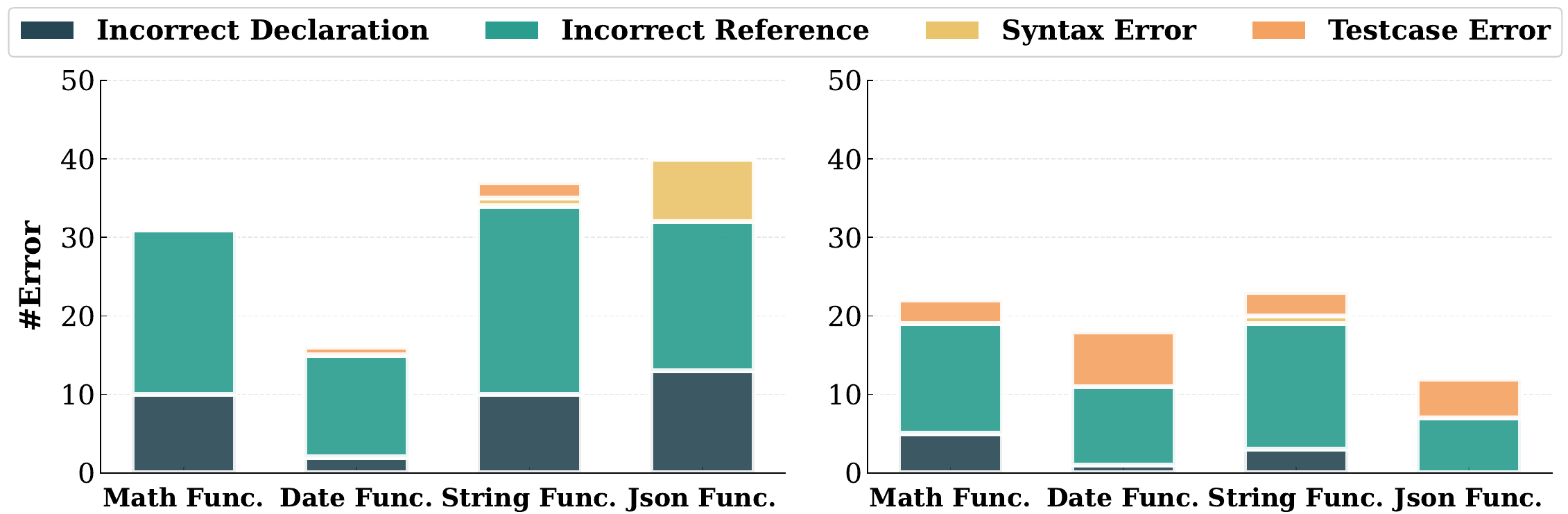}
        \caption{Error Distribution (Claude Sonnet 4.5~\cite{claudesonnet} and Claude Code~\cite{claudecode})}
    \end{subfigure}
    \caption{Native database function synthesis is hindered by dense but sparsely used references, inefficient operations, and diverse errors \normalfont{-- (a) All references versus those in native functions; (b) Operation distribution of SOTA agent-based methods (i.e., Claude Code~\cite{claudecode}); (c) Error distribution of \llm-based (Claude Sonnet 4.5~\cite{claudesonnet}) and agent-based methods (Claude Code~\cite{claudecode}).}}
    \vspace{-.65cm}
    \label{fig:pilot}
\end{figure}

\noindent \textbf{O1. Databases contain a large number of references due to their inherent complexity, while native functions depend on a small set of relevant references for their own implementation.}
We analyze the total number of references across the entire database repository and compare them with the references directly utilized by native functions.
As shown in Figure~\ref{fig:pilot} (a), we notice that although database repositories contain a vast number of file and function references, native functions depend on only a limited portion of them.
Specifically, the average number of references per file in the three database repositories is 2619.56, 1594.65, and 872.32, respectively, while the corresponding averages for a single function unit are only 13.73, 47.58, and 33.11.
This high reference density underscores the need for effective methods to accurately identify the correct reference functions during the synthesis process.


\noindent \textbf{O2. Current agent-based frameworks employ a generic design with limited focus on database-specific operations, leading to inefficient exploration and synthesis.}
Most existing code synthesis frameworks, including state-of-the-art agent-based approaches, are general-purpose systems designed for diverse code repositories~\cite{swebench}.
However, they often overlook the structural characteristics of database projects (e.g., function definitions in files such as \emph{pg\_proc.dat} in PostgreSQL), which hinder the efficient synthesis of database native functions.
As shown in Figure~\ref{fig:pilot} (b), Claude Code~\cite{claudecode} spends most of its time on file-level searches (63.70\% on \emph{Search Repo} and \emph{Read File}) instead of code generation tasks (4.95\% on \emph{Update File}), which implement the function logic.
This imbalance highlights that current frameworks focus too much on repository traversal, neglecting code construction and validation.
Therefore, there is a need for database-aware synthesis strategies that can efficiently navigate repository structures while maintaining a balanced focus on both code generation and correctness.

\begin{figure*}[!t]
  \centering
  \includegraphics[width=\linewidth]{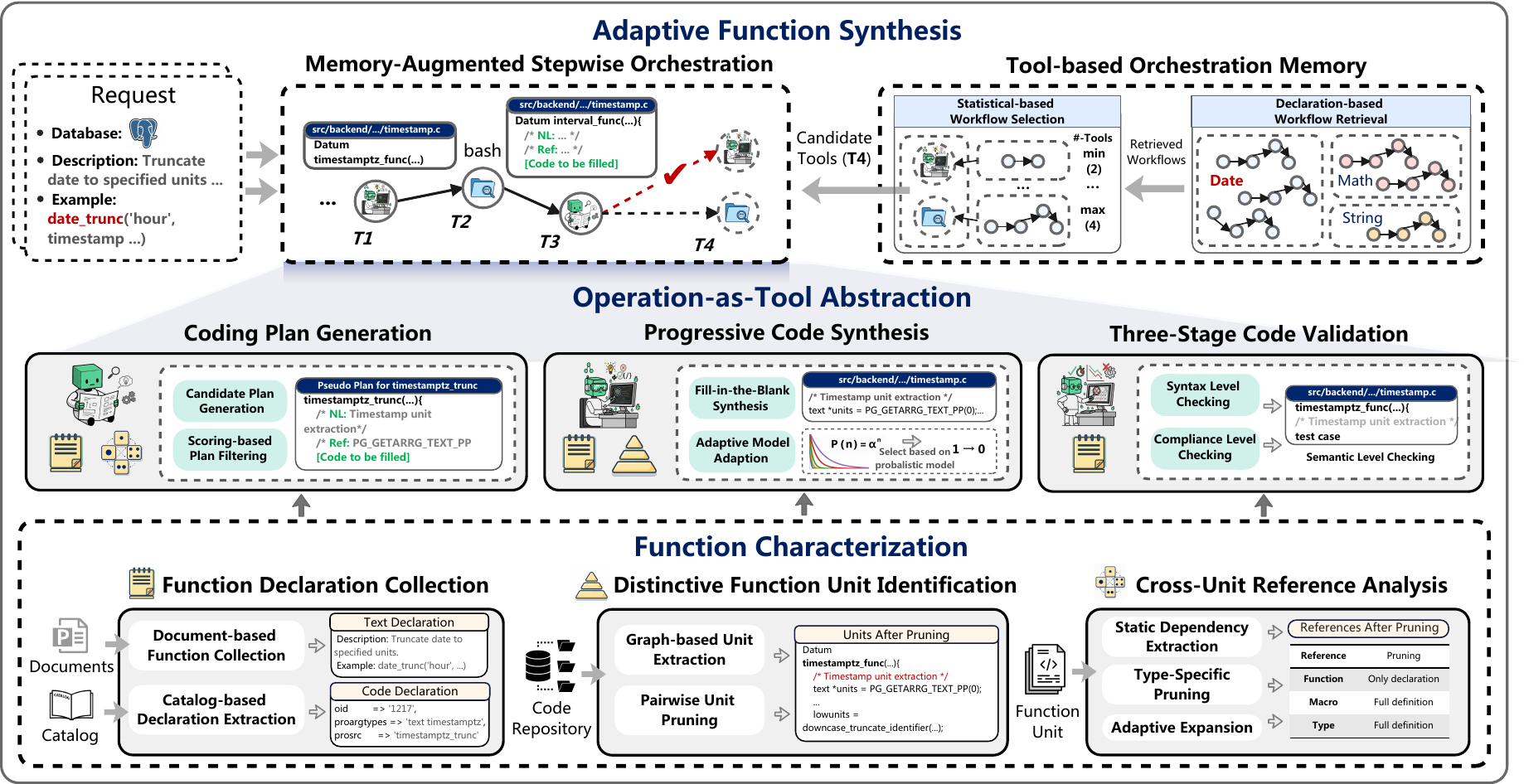}
  \vspace{-.35cm}
  \caption{Technique Overview of \oursys.}
  \label{fig:overview}
  \vspace{-.35cm}
\end{figure*}

\noindent \textbf{O3. Existing synthesis methods exhibit various errors, with declaration errors being the most frequent in synthesizing database functions.}
We analyze the error types and frequencies produced by an advanced \llm (Claude Sonnet 4.5~\cite{claudesonnet}) and an agent-based framework (Claude Code~\cite{claudecode}) when synthesizing PostgreSQL functions.
As shown in Figure~\ref{fig:pilot} (c), both methods, especially the advanced \llm, produce many declaration-related errors.
These include redeclaration within the same file (e.g., redefining \emph{int4div} in \emph{div}) and incorrect function references with mismatched arguments (e.g., too few arguments to \emph{pg\_regcomp} in \emph{regexp\_matches}).
On average, declaration errors account for 81.76\% of all errors in both methods.
We also observe test-case errors in which generated functions fail to produce the expected outputs.
These results highlight the need for improved module designs to reduce declaration issues, unresolved references, and test case failures during synthesis.





\section{\oursys~Overview}
\label{sec:overview}







In this section, we introduce the system overview (see Figure~\ref{fig:overview}).
In contrast to general-purpose code synthesis frameworks~\cite{claudecode, trae}, \oursys~is designed with explicit considerations of database codebase characteristics, enabling robust handling of their unique structural organizations (e.g., the strict mapping between SQL-level catalog definitions and their underlying implementation units).






\hi{Function Characterization.}
This module is designed around how database functions are implemented, forming structural templates and constraints that capture the internal organization to guide synthesis.
Taking the database codebase as input, we introduce automatic strategies to extract key metadata as output (defined in Definition~\ref{def:databasenativefunction}).
\textit{(1) Function Declaration Collection.} Function declarations are useful (e.g., function name, arguments) for code reuse (e.g., referenced units across similar functions) during synthesis.
To address the SQL-level declaration of database functions, we take official documentation and system catalogs (e.g., \emph{pg\_proc}) as input and collect declarations in two ways: by parsing official documentation to retrieve structured function data and by querying system catalogs (e.g., \emph{pg\_proc} in PostgreSQL) to gather authoritative declarations directly from the database engine; 
\textit{(2) Distinctive Function-Unit Identification.}
To address the database-specific mappings from a single SQL function to multiple internal units (e.g., 4 date types for the PostgreSQL date\_trunc() function), 
we adopt a graph-based analysis strategy.
Using the function name as input, we identify reusable units by locating potential registration points via keyword matching, then traversing code paths in the dependency graph (e.g., constructed by static analysis tools~\cite{understand}) to collect related units, and distinguishing shared auxiliary units (reusable as templates) from distinctive ones through pairwise comparisons; 
\textit{(3) Cross-Unit Reference Analysis.} To address the complex cross-unit
dependencies within the database codebase and reuse these references for accurate unit synthesis, we propose a context-aware method.
Taking the initial dependency graph as input, we first collect dependencies and then refine them using category-specific rules to retain only essential content (e.g., keep declarations, omit full method bodies), which can later be restored during synthesis.

\hi{Function Synthesis Operations.}
This module builds on the extracted function characters and includes specialized synthesis operations to ensure database correctness.
It involves three operations: planning, coding, and testing for native function synthesis.
\textit{(1) Pseudo-based Coding Plan Generation.}
To guide native function synthesis that follows registration rules and internal structures, we propose pseudo-plan generation to create skeletons.
Taking function metadata and database hierarchy as input, we identify relevant reference units via meta-information matching from similar functions.
Then, we instruct \llms to generate multiple candidate plans (i.e., structured code skeletons) using these units, with placeholders for variables, function bodies, and return expressions.
A score-based pruning mechanism ranks these candidates to output the optimal plan based on database code brevity and convention adherence;
\textit{(2) Progressive Code Synthesis.}
We model function synthesis as a fill-in-the-blank problem, where each blank corresponds to a required code unit.
A two-step strategy is employed: (a) a template instantiation mechanism retrieves partially matched templates based on declaration types; (b) in cases of failures, a probabilistic adaptation module controls the synthesis, allowing either blank-filling or generation from scratch to output the synthesized code;
\textit{(3) Three-Stage Code Validation.}
To ensure the compilation and correct database execution of synthesized functions, we introduce a three-tier progressive validation framework that takes the synthesized code as input: \emph{(a) Syntax validation} uses grammar parsers (e.g., ANTLR) to ensure parseability under the target database; \emph{(b) Compliance checking} uses database-native compilation tools to verify the function adheres to internal conventions (e.g., registration consistency); \emph{(c) Semantic validation} augments the database's test suite with auto-generated test cases across input types and edge cases.

\hi{Adaptive Function Synthesis.}
To address the limitations of rigid ``coding $\rightarrow$ testing'' workflows for functions of varying complexity, we first unify diverse synthesis steps, including both common utilities (e.g., \emph{bash}) and \llm-based tools, by abstracting them into a set of callable tools.
Then, taking the synthesis task as input, we develop a hybrid orchestration strategy that combines an \llm-based controller, which adaptively determines the next {tool} through context-aware reasoning, with a global trajectory memory that accumulates historical workflows to guide and accelerate synthesis {with similar functions} to output the final functions.

\section{Function Characterization}
\label{sec:offline}
In this section, we introduce the \emph{Function Characterization} module.
Unlike general-purpose agents that rely on inefficient file search~\cite{qwencode, claudecode}, this module performs deterministic graph-based analysis to decompose SQL functions into essential structural templates rather than raw snippets.
It is designed to address three challenges inherent to database kernels.
First, constructing a concise and informative overview of a function is non-trivial, as its definition is distributed across heterogeneous textual and code-level declarations.
Second, identifying the essential function units requires understanding the database's internal architectural conventions (e.g., type-specific components and modular execution pipelines).
Third, these function units often exhibit structural reuse and cross-unit dependencies (e.g., database-specific macros) to facilitate consistent implementations across similar functions.

\subsection{Function Declaration Collection}
\label{subsec:declaration}

To effectively represent functions with key information from the SQL keyword, we design a dual-source strategy to extract both the textual and the code-style function declarations $f_{dec}$ systematically.

\hi{Document-based Function Collection.}
To capture the SQL-level semantics of native functions, we automatically parse and analyze official database documentation to collect textual function declarations.
We identify sections describing native functions by examining the hierarchical document structure and employ automated scripts to parse and normalize the extracted content into a unified JSON format containing function-related information.
For instance, from PostgreSQL's ``\emph{Chapter 9. Functions and Operators}'', we extract the function \emph{date\_trunc}, described as ``\emph{truncate timestamp to specified precision}'', with examples such as \emph{date\_trunc(`hour', timestamp `2001-02-16 20:38:40')} $\rightarrow$ \emph{2001-02-16 20:00:00}.


\begin{sloppypar}
\hi{Catalog-based Declaration Extraction.}
To accurately capture implementation-level specifications, we extract definitive code-level declarations from the internal system catalogs and registration files (e.g., the \emph{pg\_proc} table and the \emph{pg\_proc.dat} file in PostgreSQL).
We query and consolidate catalog entries, systematically retrieving key attributes such as input arguments and return types, and normalize them into the same JSON format.
For example, querying PostgreSQL's system catalog for \emph{date\_trunc} returns an entry indicating two input arguments (\emph{proargtypes = text timestamptz}) and a result type of \emph{timestamptz} (\emph{prorettype = timestamptz}).
\end{sloppypar}


\begin{figure}[!t]
  \centering
  \includegraphics[width=.8\linewidth]{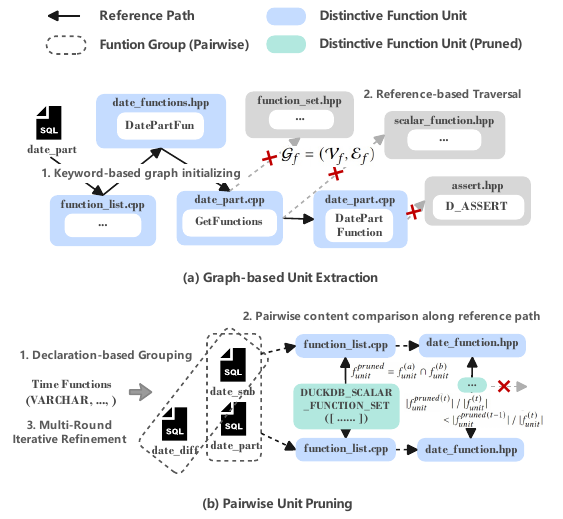}
  \vspace{-.35cm}
  \caption{Example Function Unit Identification.}
  \label{fig:unit}
  \vspace{-.5cm}
\end{figure}

\subsection{Distinctive Function Unit Identification}
\label{subsec:unit}

To follow the implicit conventions for new database function synthesis, we design two modules to identify the essential function units $f_{unit}$ that require implementation across multiple files.

\hi{Graph-based Unit Extraction.}
To uncover the underlying implementation of an encapsulated SQL function interface, we propose a graph-based unit extraction method that systematically identifies all associated function units $f_{unit}$.
For a native database function \(f\) with SQL-level declaration \(f_{dec}\), we first perform keyword-based retrieval to locate the corresponding function entry and its registration unit \(f^{reg}_{unit}\).
Starting from \(f^{reg}_{unit}\), we apply automated static analysis to construct a reference graph \(\mathcal{G}_{f} = (\mathcal{V}_{f}, \mathcal{E}_{f})\), where \(\mathcal{V}_f\) denotes the set of function units \(f_{unit}\) and \(\mathcal{E}_f\) encodes references through function invocations or class inheritance relations.
The graph expansion proceeds recursively and terminates when no additional function units are discovered or when the number of current units being referenced is larger than a predefined threshold.


As shown in Figure~\ref{fig:unit} (a), we initialize a dependency graph in DuckDB from the SQL keyword \emph{date\_part} to locate the registration entry in \emph{function\_list.cpp}.
The graph is then expanded via reference traversal to identify key functional units, such as \emph{DatePartFun} in \emph{date\_functions.hpp}, while \emph{function\_set.hpp} is excluded since the involved component \emph{ScalarFunctionSet} is a reference unit widely utilized across scalar functions.




\hi{Pairwise Unit Pruning.}  
To identify common code patterns across function units and facilitate the synthesis of similar native functions, we introduce a pairwise unit pruning strategy that extracts pruned fixed patterns from the set of extracted function units \(\{f_{unit}\}\).

\emph{(1) Grouping by Function Declarations:}
To facilitate the identification of pruned implementation patterns, the extracted function units ${f_{unit}}$ are first grouped according to their SQL-level declarations $f_{dec}$, considering input-output argument types and functional categories.
For example, the functions \emph{timestamptz\_part} and \emph{extract\_timestamptz} are grouped together because they belong to the same datetime category and accept timestamp arguments (Figure~\ref{fig:unit}).

\emph{(2) Pairwise Content Comparison:}  
To effectively derive candidate function units with common patterns, units within each group are randomly selected in pairs and compared along their invocation paths in the reference graphs $\mathcal{G}_{f} = (\mathcal{V}_{f}, \mathcal{E}_{f})$.
For a pair of units \(f_{unit}^{(a)}, f_{unit}^{(b)}\), identifiers such as variable names are first abstracted to normalize the code.
The pruned code blocks are then extracted as \(
f^{pruned}_{unit} = f_{unit}^{(a)} \cap f_{unit}^{(b)}
\) based on the exact keyword matching.
Non-pruned blocks are replaced with placeholders to represent variability across implementations.

\emph{(3) Multi-Round Refinement:}  
To robustly determine representative pruned units $f^{pruned}_{unit}$, multiple rounds of pairwise comparison are performed within each group, bounded by the group size.
Each round $t$ stops when the proportion of pruned components along the code reference paths decreases compared to the previous round $t-1$, i.e., \(|f_{unit}^{pruned (t)}|\ /\ |f^{(t)}_{unit}| < |f_{unit}^{pruned (t-1)}|\ /\ |f^{(t)}_{unit}| \).  

\begin{sloppypar}
As shown in Figure~\ref{fig:unit} (b), we first classify \emph{date\_sub()} function based on its declaration of \emph{Time Functions}.
We then randomly select \emph{date\_sub()} and \emph{date\_part()} from the same group for pairwise comparison.
Along their graph reference paths, we prune units to derive \emph{DUCKDB\_SCALAR\_FUNCTION\_SET(...)} in \emph{function\_list.cpp}, terminating at \emph{date\_functions.hpp} when pruned units become significantly smaller.
Multi-round comparisons with other function pairs, such as \emph{date\_sub()} and \emph{date\_diff()}, are conducted to robustly extract diverse pruned units.
Finally, the top-$k$ most frequently occurring pruned units $f^{pruned}_{unit}$ across comparisons are selected as representative pruned function units, capturing stable and reusable structures for the fill-in-the-blank synthesis model in Section~\ref{subsec:coding}.
\end{sloppypar}

\subsection{Cross-Unit Reference Analysis}
\label{subsec:connection}


To systematically capture dependencies among function units within complex database repositories, we propose cross-unit reference analysis to extract reference units \(f^{ref}_{unit}\) for each native function \(f\).

\emph{(1) Static Dependency Extraction:}  
Given the reference graph \(\mathcal{G}_{f} = (\mathcal{V}_{f}, \mathcal{E}_{f})\), we iteratively perform automated static analysis to identify the reference units \(\{f^{ref}_{unit}\}\) for each function unit \(f_{unit} \in \mathcal{V}_{f}\) along the graph paths, including assertion references \emph{D\_ASSERT} and executor references \emph{BinaryExecutor} in \emph{DatePartFunction}.

\emph{(2) Type-Specific Pruning:}  
To address the redundancy or excessive details introduced by directly including all reference units (e.g., complete class definitions), we apply predefined type-specific pruning rules that retain only the essential contextual information while preserving core structural elements.
For example, the \emph{ScalarFunctionSet} class retains its declaration list, while detailed method implementations, such as \emph{GetFunctionByArguments}, are omitted.

\emph{(3) Adaptive Expansion:}  
The resulting reference set provides an initial representation of \(f^{ref}_{unit}\), which can be incrementally expanded during function synthesis. Additional references can be added on demand by using automated static analysis tools to retrieve the full content of reference units, such as the complete implementation of the \emph{GetFunctionByArguments} method.

\section{Function Code Synthesis Operations}
\label{sec:synthesis}


Existing code synthesis frameworks~\cite{claudecode, qwencode, trae} have limitations in utilizing the characterized database function information.
First, they rely on \llms to directly generate code using bash commands and file-related tools, which are difficult to identify complex function relations, such as referenced macros and internal units in existing functions (Section~\ref{subsec:connection}).
Second, they generate code from scratch, whereas many database functions are large and complex (e.g., 6,235 lines across 225 functions for \emph{data\_trunc()}) and could be partially reused to reduce synthesis effort (Section~\ref{subsec:unit}).
Third, they depend on \llm's internal knowledge for validation, but database functions often take diverse arguments whose edge cases require careful handling to ensure test coverage.
To address these, we enhance database function synthesis with three key \llm-enhanced operations.

\subsection{Pseudo-based Coding Plan Generation}
\label{subsec:planning}

As illustrated in Figure~\ref{fig:intro}, human developers invest significant effort to map a SQL-level function to its internal units and determine their implementation logic and locations across files via repository exploration.
Rather than relying on invisible \llm reasoning~\cite{react, sweagent}, \oursys~defers code synthesis to an informative plan.
We generate skeletons based on registered function units (for SQL-level functionality) and internal database hierarchy information.
Each plan explicitly comprises: (1) the decomposed function units, and (2) the referenced units within the database.







\begin{figure}[!t]
  \centering
  \includegraphics[width=.85\linewidth]{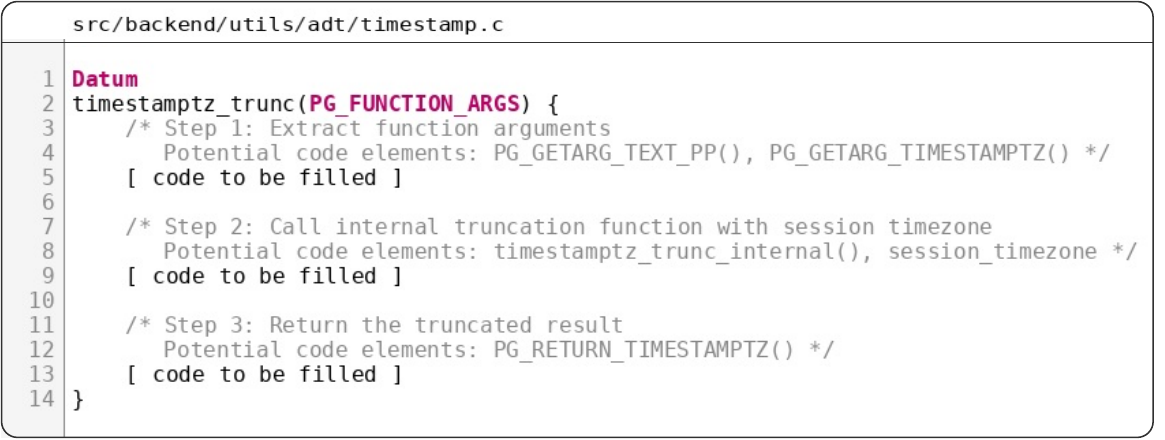}
  \caption{Example Pseudo-based Coding Plan.}
  \label{fig:plan}
  \vspace{-.35cm}
\end{figure}




\subsection*{5.1.1 Candidate Plan Generation}


To identify relevant function units for an informative plan, we adopt a metadata-based matching strategy. Specifically, we first collect potential referenced function units from all the existing functions with the same function category.
The referenced function units are then filtered to remove duplicates and grouped by unit type.

\begin{sloppypar}
Considering the diverse processing logic required for different input types and the frequent reuse of existing units in database native functions, we define and instruct \llm to generate the candidate coding plans with:
(1) the decomposed function units for each input type, as well as the structured processing logic within each unit, and
(2) the potential referenced units in the database that support the implementation of each code block in the decomposed function units.
As shown in Figure~\ref{fig:plan}, the coding plan enumerates the required function units along with their file paths (e.g., \emph{Datum timestamptz\_trunc(PG\_FUNCTION\_ARGS)} in \emph{src/backend/utils/adt/timestamp.c}). 
Furthermore, each unit is decomposed into a sequence of logical code blocks with certain functionality.
For each block, the \llm-generated plan includes a natural language description (e.g., \emph{Step 1: Extract function arguments}) and a list of potential referenced units (e.g., \emph{PG\_GETARG\_TEXT\_PP()}).
\end{sloppypar}


While these plans outline essential details for code implementations, they might incur errors in plan generation, such as invalid reference units or incomplete processing logic.
For instance, when synthesizing PostgreSQL's \emph{date\_trunc()} function, an incorrect plan might use a non-existent macro (e.g., \emph{PG\_GETARG\_TXT()}) or omit the essential processing branches for the \emph{interval} input type.
To further improve the quality of generated plans, we employ an ensemble strategy that generates multiple pseudo-plans in parallel and refines them through a scoring-based filtering process.

\hypertarget{subsubsec:plan_filtering}{}
\subsection*{5.1.2 Scoring-based Plan Filtering}

To reduce the risk that the generated low-quality plans might mislead the synthesis process, we introduce scoring-based plan filtering based on a metric-based scoring function.
Unlike existing methods that rely on \llm to produce uncertain evaluation results~\cite{qwencode}, this function deterministically assesses the generated plans, considering both faithfulness (i.e., the correctness of the listed referenced units) and simplicity (i.e., the number of code blocks).
Specifically, given $n$ implementation plans, it computes a normalized score for each plan based on three criteria.
The score for plan $k$ is computed as:
\[
R_k = \alpha \cdot N(v_1^k) + \beta \cdot N(v_2^k) + \gamma \cdot N(v_3^k)
\]
where the first two items characterize faithfulness, and the last item captures simplicity.
$v_1^k$ is the number of incorrectly referenced function units (i.e., the ones listed in ``\emph{Potential code elements}'').
$v_2^k$ is the number of incorrectly specified file locations for each function unit (e.g., the ones outside the specified database directory).
$v_3^k$ is the number of the listed function units.
$N(x)$ denotes min-max normalization $N(x) = 1 - \frac{x - \min}{\max - \min}$ (or 1.0 if $\min = \max$) to scale the value among the $n$ plans.
$\alpha$, $\beta$, and $\gamma$ are the weight coefficients to trade between criteria, which are $0.4$, $0.4$, and $0.2$, respectively.

We remove incorrect function units and referenced function units from the plan while computing this score, and filter out plans whose score is lower than a predefined threshold (e.g., $0.5$).





\subsection{Progressive Code Synthesis}
\label{subsec:coding}

To alleviate the risk of introducing errors during function synthesis, we propose progressive code synthesis that formulates function synthesis as a fill-in-the-blank task.
Unlike traditional methods that require \llms to synthesize the entire functions from scratch, we employ a probabilistic, self-correcting framework utilizing templates and scored pseudo-plans.
This approach progressively refines code by focusing on key function units and is uniquely capable of triggering a semantic rollback (i.e., switching from template-based to from-scratch synthesis) based on validation feedback.



\noindent \textbf{Fill-in-the-Blank Synthesis Model.}
Building on the progressive synthesis paradigm, we introduce a fill-in-the-blank synthesis model that directs \llm's attention to critical function units within native database functions.
Given the metadata (e.g., function category) specified in a function declaration, we first retrieve native functions that share the same metadata value.
From these retrieved functions, we extract the top-$k$ representative function units with placeholders in Section~\ref{subsec:unit} which contain the most relevant implementation logic that can be reused.
For example, in DuckDB, the \llm only needs to generate the distinctive function units for ten type-specific functions and add them to \emph{DatePartFun::GetFunctions}, without needing to implement all function units, such as those for registrations in \emph{DatePartFun}.
To further enhance synthesis reliability, we invoke \llm to generate multiple candidate function units in parallel.
A self-consistency strategy is then applied to merge these candidates by selecting units that appear most frequently across the generated results, ensuring coherent and high-quality synthesis.

\begin{figure}[!t]
  \centering
  \includegraphics[width=.8\linewidth]{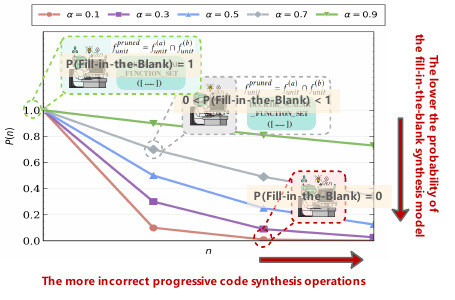}
  \vspace{-.35cm}
  \caption{Synthesis Model Adaption Example.}
  \label{fig:model}
  \vspace{-.6cm}
\end{figure}


\noindent \textbf{\hypertarget{subsubsec:model_adaption}{Synthesis Model Adaption}.}
To address the issue of incorrectly identified function units in Section~\ref{subsec:unit} and enhance overall synthesis flexibility, we introduce an automatic adaptation mechanism that dynamically regulates the application of the fill-in-the-blank model based on historical synthesis.
As shown in Figure~\ref{fig:model}, we formulate the adoption of the fill-in-the-blank model as a probabilistic problem, where the initial probability of using the fill-in-the-blank model is set to \(P_0 = 1\).
This probability decays exponentially with each failed synthesis attempt according to: \( P(n) = \alpha ^n\), where \( n \) denotes the number of code synthesis failures and \( \alpha \) (\( 0 < \alpha < 1 \)) is the decay factor that determines the probability reduction rate.
A smaller value of \( \alpha \) corresponds to a faster decrease in the model adoption probability.
Once the probability reaches zero (i.e., \(P_0 = 0\)), the \llm shifts to a fully exploratory synthesis mode, autonomously identifying and implementing all required function units from scratch.
This transition serves as a rollback mechanism for handling errors such as incorrect coding plans.
For example, when a plan indicates a non-existent macro or an incorrect file path, \oursys~can entirely ignore the flawed plan after several failed attempts and eventually implement the logic from scratch by searching the repository.



\subsection{Three-Stage Code Validation}
\label{subsec:testing}

Existing approaches rely on the \emph{bash} tools and \llm-generated \emph{make} commands to validate function codes, which might introduce instability in the validation process.
Unlike these methods that use generic syntax checks or static external tests~\cite{swebench}, we propose a progressive validation strategy that delivers timely and reliable feedback by assessing generated function units at three hierarchical levels, from single-file correctness to multi-file integration.
This strategy integrates with the databases, employs kernel-level compliance checks, and an \llm-augmented test generator to ensure comprehensive correctness.
\emph{(1) Syntax Level:}
This level performs basic syntax checks within a single file to ensure variables are properly declared and referenced, using language-specific parsers like ANTLR~\cite{antlr};
\emph{(2) Compliance Level:}
This level verifies adherence to database-specific conventions across multiple files by customizing compliance tools with commands (e.g., \emph{make install} in PostgreSQL), validating correct integration and build compliance within a specified timeout;
\emph{(3) Semantic Level:}
This level evaluates whether generated functions produce expected outputs on representative test cases.
We use \llm to generate SQLs to assess runtime behavior, maximizing coverage with three contextual sources:
(a) Expertise instructions to guide \llm on error-prone areas like input types;
(b) Existing database test suites for reusable examples and format conventions;
(c) Decomposed internal code blocks to generate test cases relevant to the new implementation.

\section{From Static Function Synthesis to Adaptive Tool Orchestration}
\label{sec:orchestration}

\begin{sloppypar}
Conventional synthesis workflows typically adopt pre-defined pipelines with fixed stages (e.g., \emph{coding} $\rightarrow$ \emph{testing})~\cite{trae}.
However, database native functions exhibit diverse synthesis complexities.
For example, PostgreSQL implements mathematical functions (e.g., \emph{sin}) through concise wrappers around the standard library, while others require the implementations of specific processing logic.
To overcome the limited flexibility of pre-defined workflows in handling function-specific synthesis, we shift from static synthesis workflows to an adaptive tool orchestration framework.
Unlike rigid pipelines with fixed workflows~\cite{metagpt, opendevin}, we employ a memory-augmented strategy that dynamically sequences operations using retrieved similar trajectories and synthesis context.
\end{sloppypar}

\subsection{Operation-as-Tool Abstraction}
\label{subsec:tool}

To seamlessly integrate synthesis operations with common tools (e.g., file search) into a unified workflow, we introduce operation-as-tool abstraction, which encapsulates each synthesis operation as a callable tool $\mathcal{O}$.
Each tool is implemented with a unified interface that standardizes invocation and result handling across different operations.
The tool interface comprises three modules.

\begin{itemize}[leftmargin=10pt,topsep=2pt]
    \item \emph{(1) Metadata Module:} Define the tool's name, the arguments, and the functionality descriptions.
    
    \item \emph{(2) Core Logic Module:} Implement the designed operations via the processing logic module, such as coding plan generation.
    
    \item \emph{(3) Routing Module:} Forward the input to the processing logic module and return the outputs in a standardized format that can be directly fed into the context of \llm.
\end{itemize}

For example, we encapsulate the coding plan operation as a tool with: (1) the \emph{metadata module} including the name called ``plan\_agent'', the input argument including ``plan\_num'' for the expected plan number, and the illustration describing ``Generate pseudo-based plans to outline and instruct synthesis'';
(2) the \emph{core logic module} to instruct \llm to generate multiple pseudo-based plans, and filter these plans based on pre-defined scoring functions;
(3) the \emph{routing module} to return the plans in JSON contents to input to \llm.

We integrate a hybrid tool set with:
(1) \emph{Common utilities} including bash command interfaces and static analysis tools for function characterization;
(2) \emph{\llm-based operations} covering proposed synthesis operations (e.g., pseudo-based plan generation) as well as \llm-optimization strategies such as majority-vote sampling and post-synthesis self-reflection.


\begin{algorithm}[!t]
\caption{Tool-based Stepwise Function Synthesis}
\label{algo:synthesis}
\KwIn{Function Declaration $f_{dec}$, Synthesis Tool Set $\mathcal{O}$, Synthesis Memory Pool $\mathcal{M}$}
\KwOut{Synthesized Function Units $f_{unit}$}

\BlankLine
\gray{\textsf{/ * 1. Get trajectory of similar functions as reference */}} \\
Retrieve items $\mathcal{M}'_{\text{ref}} = \{ m \in \mathcal{M} \mid \text{category}(m) = \text{category}(f_{dec}) \wedge \text{count\_op}(m) \in \{\min, \text{median}, \max\}\}$ \label{algo:retrieve}

\BlankLine
\gray{\textsf{/ * 2. Orchestrate operation tool on-the-fly */}} \\
Initialize function unit $f_{unit} \leftarrow \emptyset$, trajectory $\mathcal{C} \leftarrow \emptyset$, tool $op \leftarrow {op_{coding}} \in \mathcal{O} = \{{op}_{plan}, {op}_{code}, ..., {op}_{validation} \}$ \\
\While{True}{
      Update $f_{unit} \leftarrow \text{execute}{(op)}$ and update $\mathcal{C} \leftarrow \mathcal{C} \cup {op}$ \\
    Determine next tool $op' \leftarrow \text{LLM}(\mathcal{C}, \mathcal{M}'_{ref})$ \\ \label{algo:tool}
    \If{$op' = {op}_{stop}$}{
    Generate trajectory summary $s \leftarrow \text{LLM}(\mathcal{C})$ \\
    \textbf{break}
    }
    \Else{$op \leftarrow op'$}
}

\BlankLine
\gray{\textsf{/ * 3. Save trajectory and return synthesized units */}} \\
Store $(f_{dec}, \mathcal{C}, s)$ into memory pool $\mathcal{M}$

\BlankLine
\Return $f_{unit}$ \label{algo:result}
\end{algorithm}

\subsection{Tool-based Stepwise Function Synthesis}
\label{subsec:orchestration}

To flexibly handle the synthesis of different functions, we introduce a tool-based stepwise synthesis framework that decomposes full implementations into the adaptive invocation of various tools.

\noindent \textbf{Tool-based Orchestration Memory.}
To move beyond the limitations of \llm-only orchestration, we develop a tool-based orchestration memory that provides external and structured guidance for adaptive synthesis.
The memory pool $\mathcal{M} = \{f_{dec}, \mathcal{C}, s\}$ is built to capture orchestrations with:
(1) the function metadata, such as the category in the function declaration $f_{dec}$;
(2) the tool invocation trajectory $\mathcal{C}$, which captures the tool sequence and associated statistics (e.g., number of each invoked tool); 
(3) a concise synthesis summary $s$ describing the overall orchestration process as generated by \llm.
To ensure the trajectory pool remains compact and representative, we implement a distribution-aware quality control mechanism.
Trajectories are organized by function category (e.g., \textit{math} and \textit{date}) to enable efficient retrieval of orchestration patterns for similar operations. A new trajectory is inserted only when it provides new statistical evidence for its category.
Specifically, insertion occurs when a trajectory's tool-invocation profile (i.e., the number and types of invoked tools) updates the maintained statistics of the pool (e.g., minimum, median, maximum tool usage).

\noindent \textbf{Memory-Augmented Stepwise Orchestration.}
As illustrated in Algorithm~\ref{algo:synthesis}, the synthesis begins by retrieving relevant entries from $\mathcal{M}$ via metadata matching (e.g., same category).
To ensure balanced reference coverage, the entries with minimum, median, and maximum tool counts are selected to form the orchestration reference $\mathcal{M}^{\prime}_{ref}$ (\textbf{line~\ref{algo:retrieve}}).
Starting from the \llm-based tool $op$ (e.g., code synthesis operation), we invoke the selected tools to iteratively update the synthesized function unit $f_{unit}$ and the orchestration trajectory $\mathcal{C}$.
\llm dynamically determines the next tool $op^{\prime}$ (e.g., pseudo-plan generation for initial implementation outlining or a new plan generation to fix errors identified by code validation) based on the evolving synthesis context (\textbf{line~\ref{algo:tool}}).
If the determined next tool $op_{\prime}$ aims to stop the orchestration process, we invoke \llm to produce the final trajectory summary $s$ and store the complete orchestration information $(f_{dec}, \mathcal{C}, s)$ into the memory pool $\mathcal{M}$, and return the synthesized function units $f_{units}$ (\textbf{line~\ref{algo:result}}).




\section{Experiments}
\label{sec:exp}


We conduct experiments to validate the effectiveness of \oursys.

\subsection{Experiment Settings}
\label{subsec:setup}


\begin{sloppypar}
\hi{Tested Databases.} We synthesize native functions over three mainstream databases:
(1) SQLite: A lightweight database engine implemented in C;
(2) PostgreSQL: A full-featured, standards-compliant object-relational database written in C;
(3) DuckDB: A high-performance in-process database for analytical processing written in C++.
These databases coherently vary in complexity, where SQLite is more lightweight than the other two databases.
\end{sloppypar}

\hi{Tested Functions.} We synthesize functions in two types:
(1) those with ground-truth implementations in the official repositories, and (2) those currently absent from these repositories.
To ensure representative coverage across diverse operations and complexity, we first classify functions based on categories defined in the official documents (e.g., date and time).
We then select functions from each category based on their implementation complexity (e.g., the code length and the number of reference units).
In total, we test 75, 145, and 128 functions on SQLite, PostgreSQL, and DuckDB, respectively.
For functions absent from a target database (e.g., SQLite), we collect distinct function declarations from the other two databases (e.g., PostgreSQL and DuckDB) as synthesis inputs.
The complete list of tested functions is included in our \href{https://github.com/weAIDB/DBCooker}{\gray{[\underline{artifact}]}}.



\noindent \textbf{Evaluation Methods.}
We assess three types of methods.
(1) \llm-based: We evaluate advanced \llms including GPT-5~\cite{gpt5}, Claude Opus 4.1~\cite{claudeopus}, Claude Sonnet 4.5~\cite{claudesonnet}, and Qwen3 Coder Plus~\cite{qwen3coder}.
Each model is prompted to specify the exact file path for the generated code, which is then integrated into the repository using empirical placement rules (e.g., inserting function units before the \emph{aBuiltinFunc[]} definition in SQLite).
(2) RAG-based: We enhance \llm-based methods with CodeRAG, which retrieves ground-truth reference function units via static analysis and provides them as the context input to \llms.
(3) Agent-based: We assess state-of-the-art coding agents, including Claude Code~\cite{claudecode}, Qwen Code~\cite{qwencode}, and TRAE~\cite{trae} (top-1 on the SWE-bench Verified leaderboard~\cite{swebench}).


\noindent \textbf{Evaluation Metrics.}
We calculate the synthesis accuracy with two metrics. 
(1) {Compliance Accuracy (\textsf{$Acc_{EXE}$}):} The ratio of the synthesized functions that successfully compile and integrate into the database (e.g., the \emph{./configure \&\& make \&\& make install} command in PostgreSQL);
(2) {Result Accuracy} (\textsf{$Acc_{RES}$):} The ratio of the synthesized functions that pass all the testcases (e.g., those under \emph{src/test/regress/sql} in PostgreSQL) and yield expected results.

\noindent \textbf{Implementation}.
Experiments are performed on a workstation with 2 Intel Xeon E5-2678 v3 CPUs (2.50 GHz), 256 GB RAM, and 4 NVIDIA RTX 4090 Ti GPUs.
The default \llm of agent-based methods is Qwen3 Coder Plus with the temperature set to 0.1.
Each synthesis has a maximum timeout of 300 seconds.
The parameters in Section~\ref{subsec:planning} and~\ref{subsec:coding} are lightweight heuristics and remain identical across databases in our experiments and require no per-database tuning.
Following the quality control mechanism in Section~\ref{subsec:orchestration}, the resulting trajectory memory pool contains 90 entries for SQLite (filtered from 550 initial trajectories across 75 functions), 174 for PostgreSQL (from 830 trajectories across 145 functions), and 154 for DuckDB (from 762 trajectories across 128 functions).

\begin{figure*}[!t]
  \centering
  \includegraphics[width=\linewidth]{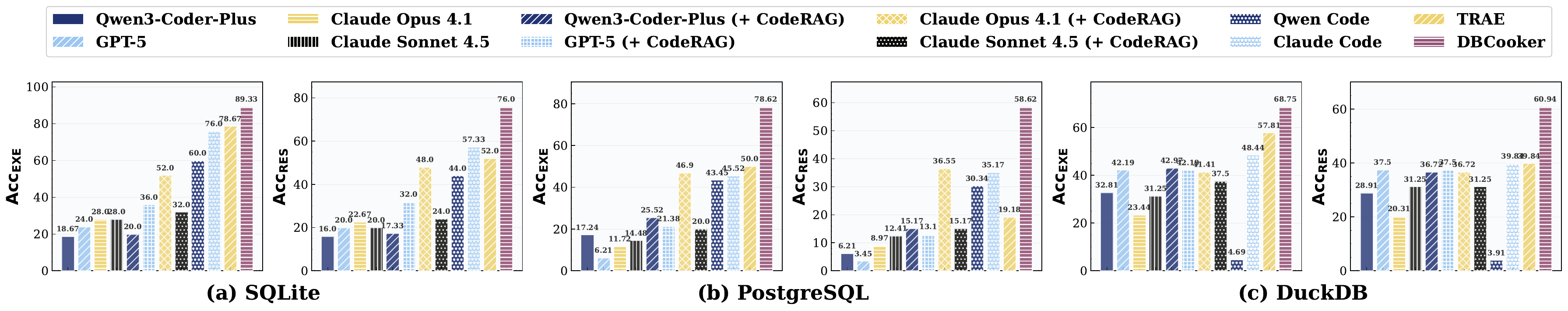}
  \vspace{-.5cm}
  \caption{Overall Code Synthesis Accuracy (\%) of Different Synthesis Methods.}
  \label{fig:overall}
  \vspace{-.35cm}
\end{figure*}

\subsection{Overall Performance}
\label{subsec:overall}

We evaluate the synthesis accuracy and analyze the error distributions of different methods across three mainstream databases.

\noindent \textbf{Synthesis Accuracy.}
According to the results in Figure~\ref{fig:overall}, we have the following two observations.
\emph{First, \oursys~outperforms all the other methods with the highest synthesis accuracy across different databases.}
Specifically, it achieves $Acc_{EXE}$ and $Acc_{RES}$ scores of 78.90\% and 65.19\%, respectively, outperforming other methods by an average margin of 124.37\% and 149.68\%, respectively.
This improvement stems from the three database-aware modules integrated in \oursys.
The \textit{function characterization} module enables \llm~to recognize system-specific specifications during synthesis, such as declaring function items in \emph{aBuiltinFunc[]} for implementations in \emph{src/func.c} of SQLite.
The \textit{adaptive orchestration} module coordinates the synthesis operations, decomposes function units, and identifies relevant reference units (e.g., adding \emph{array\_cross\_product} in \emph{ScalarFunctionSet} for functions implemented as \emph{ArrayFixedCombine<float, CrossProductOp, 3>} and \emph{ArrayFixedCombine<double, CrossProductOp, 3>}).
In contrast, \llm-based methods mostly depend on their internal knowledge, often resulting in incorrect file placements (e.g., misplacing \emph{\#include float8.h} in PostgreSQL).
Agent-based methods are inefficient due to the exploration of irrelevant parts of the repository.
For instance, Claude Code and TRAE spend excessive time performing numerous search and read operations on unrelated files.
Similarly, Qwen Code underperforms on DuckDB, repeatedly attempting full compilations with incorrect function implementations.

\begin{figure}[!t]
  \centering
  \includegraphics[width=.8\linewidth]{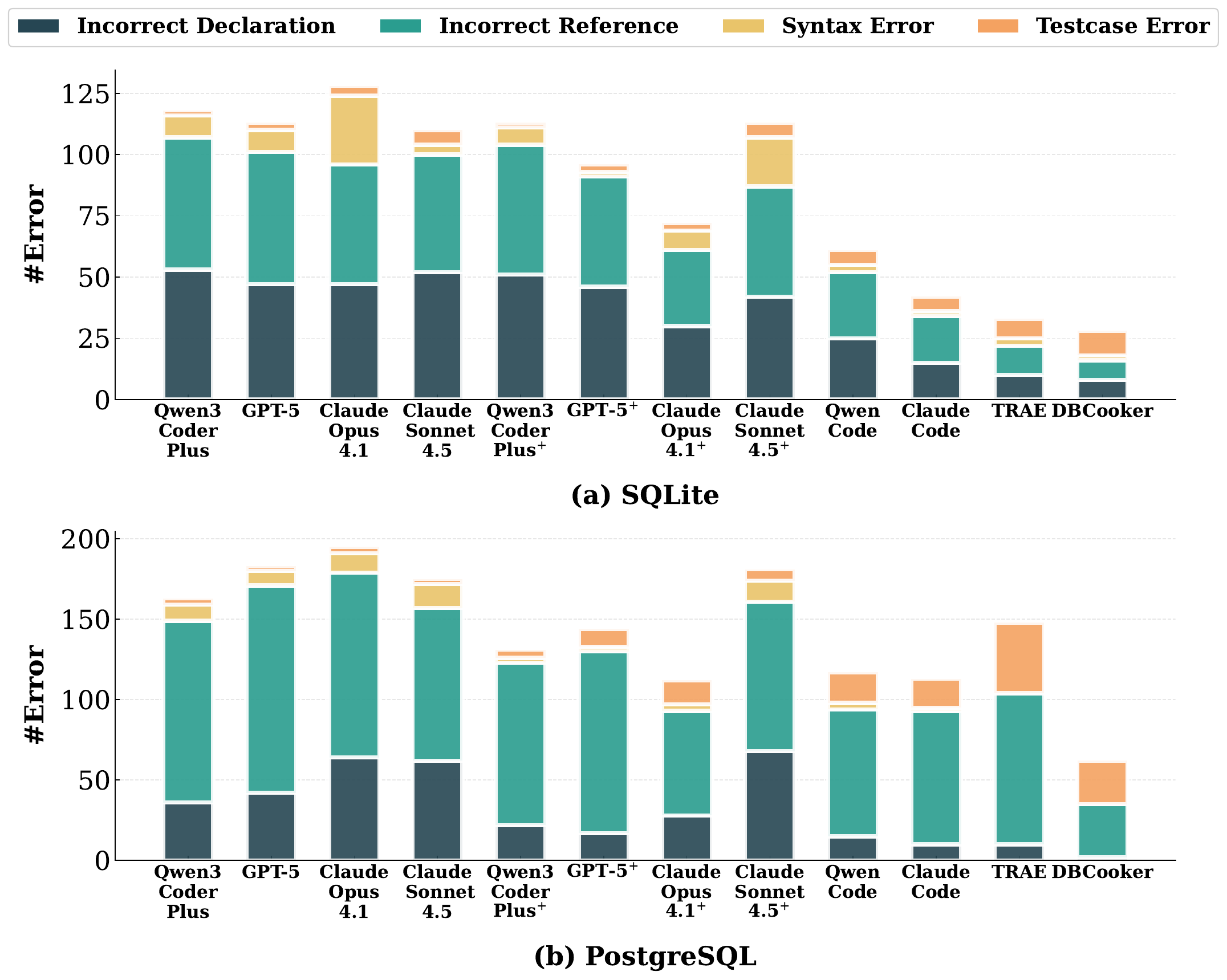}
  \vspace{-.75em}
  \caption{Error Distribution of Different Synthesis Methods ($^{+}$ indicates the usage of CodeRAG, and the codes of a native function might contain multiple errors).}
  \label{fig:error}
  \vspace{-1.75em}
\end{figure}

\emph{Second, RAG-based methods, which enrich the \llm~context with reference function units, improve the synthesis accuracy of \llm-based approaches.}
Specifically, integrating CodeRAG results in an average accuracy improvement of 50.56\% over \llm-based methods.
This improvement arises because native functions rely on multiple reference units within complex database repositories, which \llms often fail to identify on their own, despite being extensively pre-trained on the database repository.
CodeRAG addresses this by efficiently retrieving relevant function units from the large search space and providing them to the \llm, facilitating accurate integration of these units into each function's implementation (e.g., 4 structs and 18 functions for the \emph{array\_cross\_product} function in DuckDB).

\noindent \textbf{Synthesis Error Distribution.} 
We analyze the four error type distribution from different synthesis methods.
For instance, declaration-related errors, including \emph{incorrect declaration} and \emph{incorrect reference}, indicate whether a method correctly identifies declaration locations and incorporates the appropriate reference function units within the database.
As shown in Figure~\ref{fig:error}, agent-based methods produce fewer declaration-related errors, and \oursys~achieves the lowest declaration-related error.
The reduction stems from the ability of agent-based methods to actively search the repository and verify the existence of required declarations (e.g., checking whether \emph{repeat()} is registered in \emph{pg\_proc.dat}).
\oursys~further optimizes by identifying all necessary units for each function (e.g., registering \emph{repeat} in \emph{pg\_proc.dat} and implementing key units in \emph{oracle\_compat.c}) and signaling absent references such as `JsonParse' during progressive code validation.
In contrast, \llm-based methods rely mainly on internal knowledge and the input context, which often fails to capture all relevant function units, leading to missing definitions (e.g., `repeat' in PostgreSQL) or incorrect references to absent units (e.g., `JsonParse' in SQLite's \emph{json\_remove()} function).

\begin{figure}[!t]
  \centering
  \includegraphics[width=\linewidth]{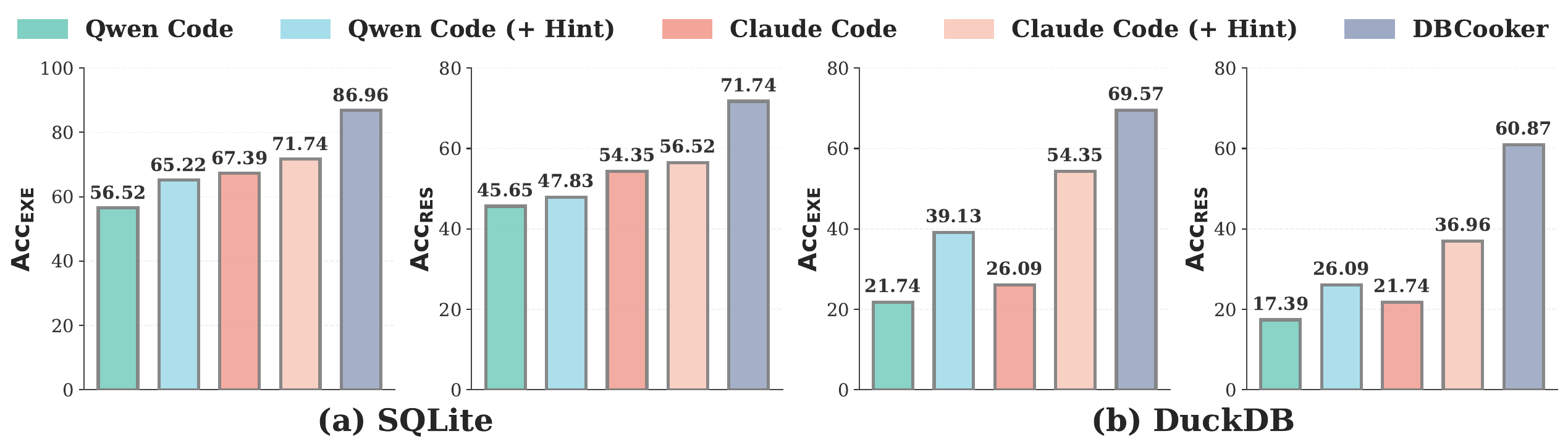}
  \vspace{-.35cm}
  \caption{Code Synthesis Accuracy (\%) with Full Context.}
  \label{fig:revision_hint}
  \vspace{-.35cm}
\end{figure}

To further isolate the errors introduced by file search, we conduct additional experiments where agentic baselines were explicitly given the full context (\textsf{+ Hint}) of correct file paths, function declarations, and reference units to estimate
their performance outside the file search problem.
As shown in Figure~\ref{fig:revision_hint}, even with this complete context, these agents (\textsf{+ Hint}) still generate code containing undefined references and type-related semantic errors (e.g., introducing numeric corruption by using \emph{int64} instead of \emph{double}), with an average 22.56\% lower accuracy than \oursys. This demonstrates that, beyond the challenge of extensive file search, failures of agentic baselines can also arise from issues such as incorrect declarations and references in their generated code, whereas \oursys~benefits from its integrated components, such as progressive validation.

\begin{sloppypar}
\noindent \textbf{Summary:}
\emph{(1) Consistent Performance Improvement.}
\oursys~outperforms both general agent-based methods (e.g., Claude Code), which struggle with complex database codebases by performing excessive searches and producing incorrect compilations, and the \textsf{+Hint} variants by 22.56\%, which effectively eliminate retrieval bottlenecks and serve as the upper bound for retrieval-based systems.
Even provided with the correct files, these methods still persist in generating code with undefined references and type-related semantic errors;
\emph{(2) Synthesis Goes Beyond Retrieval:}
These results show that database native function synthesis goes beyond retrieval.
While retrieval helps to identify the correct context (e.g., file paths and references), effective synthesis further requires enforcing database-specific correctness, such as generating type-dependent execution branches (e.g., for \emph{date\_trunc()}) and refining code under kernel-level constraints (e.g., successful compilation and passing SQL tests).
\end{sloppypar}

\subsection{Fine-Grained Analysis}
\label{subsec:fine-grained}

To further evaluate the adaptability of \oursys~across diverse functions, we perform a fine-grained analysis in two dimensions.

\begin{table*}[!t]
\caption{Code Synthesis Accuracy (\%) over Varying Function Difficulty.}
\label{tab:difficulty}
\resizebox{\linewidth}{!}{
\begin{tabular}{|l|cccccc|cccccc|cccccc|}
\hline
\multicolumn{1}{|c|}{\multirow{3}{*}{\textbf{Method}}}                           & \multicolumn{6}{c|}{\textbf{SQLite}}                                                                                                                                                                                                                   & \multicolumn{6}{c|}{\textbf{PostgreSQL}}                                                                                                                                                                                                               & \multicolumn{6}{c|}{\textbf{DuckDB}}                                                                                                                                                                                                                   \\ \cline{2-19} 
\multicolumn{1}{|c|}{}                                                           & \multicolumn{2}{c|}{\textbf{EASY}}                                                      & \multicolumn{2}{c|}{\textbf{MEDIUM}}                                                    & \multicolumn{2}{c|}{\textbf{HARD}}                                 & \multicolumn{2}{c|}{\textbf{EASY}}                                                      & \multicolumn{2}{c|}{\textbf{MEDIUM}}                                                    & \multicolumn{2}{c|}{\textbf{HARD}}                                 & \multicolumn{2}{c|}{\textbf{EASY}}                                                      & \multicolumn{2}{c|}{\textbf{MEDIUM}}                                                    & \multicolumn{2}{c|}{\textbf{HARD}}                                 \\ \cline{2-19} 
\multicolumn{1}{|c|}{}                                                           & \multicolumn{1}{c|}{\textbf{$\bm{Acc_{EXE}}$}} & \multicolumn{1}{c|}{\textbf{$\bm{Acc_{RES}}$}} & \multicolumn{1}{c|}{\textbf{$\bm{Acc_{EXE}}$}} & \multicolumn{1}{c|}{\textbf{$\bm{Acc_{RES}}$}} & \multicolumn{1}{c|}{\textbf{$\bm{Acc_{EXE}}$}} & \textbf{$\bm{Acc_{RES}}$} & \multicolumn{1}{c|}{\textbf{$\bm{Acc_{EXE}}$}} & \multicolumn{1}{c|}{\textbf{$\bm{Acc_{RES}}$}} & \multicolumn{1}{c|}{\textbf{$\bm{Acc_{EXE}}$}} & \multicolumn{1}{c|}{\textbf{$\bm{Acc_{RES}}$}} & \multicolumn{1}{c|}{\textbf{$\bm{Acc_{EXE}}$}} & \textbf{$\bm{Acc_{RES}}$} & \multicolumn{1}{c|}{\textbf{$\bm{Acc_{EXE}}$}} & \multicolumn{1}{c|}{\textbf{$\bm{Acc_{RES}}$}} & \multicolumn{1}{c|}{\textbf{$\bm{Acc_{EXE}}$}} & \multicolumn{1}{c|}{\textbf{$\bm{Acc_{RES}}$}} & \multicolumn{1}{c|}{\textbf{$\bm{Acc_{EXE}}$}} & \textbf{$\bm{Acc_{RES}}$} \\ \hline
\textbf{Qwen3-Coder-Plus}                                                        & \multicolumn{1}{c|}{22.58}                 & \multicolumn{1}{c|}{22.58}                 & \multicolumn{1}{c|}{16.67}                 & \multicolumn{1}{c|}{12.5}                  & \multicolumn{1}{c|}{15.0}                  & 10.0                  & \multicolumn{1}{c|}{12.82}                 & \multicolumn{1}{c|}{7.69}                  & \multicolumn{1}{c|}{20.45}                 & \multicolumn{1}{c|}{2.27}                  & \multicolumn{1}{c|}{26.09}                 & 8.7                   & \multicolumn{1}{c|}{41.43}                 & \multicolumn{1}{c|}{37.14}                 & \multicolumn{1}{c|}{20.45}                 & \multicolumn{1}{c|}{18.18}                 & \multicolumn{1}{c|}{28.57}                 & 21.43                 \\ \hline
\textbf{GPT-5}                                                                   & \multicolumn{1}{c|}{29.03}                 & \multicolumn{1}{c|}{29.03}                 & \multicolumn{1}{c|}{33.33}                 & \multicolumn{1}{c|}{20.83}                 & \multicolumn{1}{c|}{5.0}                   & 5.0                   & \multicolumn{1}{c|}{7.69}                  & \multicolumn{1}{c|}{6.41}                  & \multicolumn{1}{c|}{0.0}                   & \multicolumn{1}{c|}{0.0}                   & \multicolumn{1}{c|}{13.04}                 & 0.0                   & \multicolumn{1}{c|}{52.86}                 & \multicolumn{1}{c|}{48.57}                 & \multicolumn{1}{c|}{29.55}                 & \multicolumn{1}{c|}{25.0}                  & \multicolumn{1}{c|}{28.57}                 & 21.43                 \\ \hline
\textbf{Claude Opus 4.1}                                                         & \multicolumn{1}{c|}{35.48}                 & \multicolumn{1}{c|}{35.48}                 & \multicolumn{1}{c|}{33.33}                 & \multicolumn{1}{c|}{16.67}                 & \multicolumn{1}{c|}{10.0}                  & 10.0                  & \multicolumn{1}{c|}{11.54}                 & \multicolumn{1}{c|}{8.97}                  & \multicolumn{1}{c|}{6.82}                  & \multicolumn{1}{c|}{4.55}                  & \multicolumn{1}{c|}{21.74}                 & 17.39                 & \multicolumn{1}{c|}{32.86}                 & \multicolumn{1}{c|}{30.0}                  & \multicolumn{1}{c|}{11.36}                 & \multicolumn{1}{c|}{9.09}                  & \multicolumn{1}{c|}{14.29}                 & 7.14                  \\ \hline
\textbf{Claude Sonnet 4.5}                                                       & \multicolumn{1}{c|}{35.48}                 & \multicolumn{1}{c|}{32.26}                 & \multicolumn{1}{c|}{37.5}                  & \multicolumn{1}{c|}{20.83}                 & \multicolumn{1}{c|}{5.0}                   & 0.0                   & \multicolumn{1}{c|}{14.1}                  & \multicolumn{1}{c|}{10.26}                 & \multicolumn{1}{c|}{13.64}                 & \multicolumn{1}{c|}{13.64}                 & \multicolumn{1}{c|}{17.39}                 & 17.39                 & \multicolumn{1}{c|}{40.0}                  & \multicolumn{1}{c|}{37.14}                 & \multicolumn{1}{c|}{18.18}                 & \multicolumn{1}{c|}{18.18}                 & \multicolumn{1}{c|}{28.57}                 & 21.43                 \\ \hline
\textbf{\begin{tabular}[c]{@{}l@{}}Qwen3-Coder-Plus\\ (+ CodeRAG)\end{tabular}}  & \multicolumn{1}{c|}{32.26}                 & \multicolumn{1}{c|}{32.26}                 & \multicolumn{1}{c|}{16.67}                 & \multicolumn{1}{c|}{8.33}                  & \multicolumn{1}{c|}{5.0}                   & 5.0                   & \multicolumn{1}{c|}{26.92}                 & \multicolumn{1}{c|}{20.51}                 & \multicolumn{1}{c|}{25.0}                  & \multicolumn{1}{c|}{6.82}                  & \multicolumn{1}{c|}{21.74}                 & 13.04                 & \multicolumn{1}{c|}{52.86}                 & \multicolumn{1}{c|}{47.14}                 & \multicolumn{1}{c|}{34.09}                 & \multicolumn{1}{c|}{27.27}                 & \multicolumn{1}{c|}{21.43}                 & 14.29                 \\ \hline
\textbf{\begin{tabular}[c]{@{}l@{}}GPT-5\\ (+ CodeRAG)\end{tabular}}             & \multicolumn{1}{c|}{54.84}                 & \multicolumn{1}{c|}{51.61}                 & \multicolumn{1}{c|}{33.33}                 & \multicolumn{1}{c|}{29.17}                 & \multicolumn{1}{c|}{10.0}                  & 5.0                   & \multicolumn{1}{c|}{24.36}                 & \multicolumn{1}{c|}{21.79}                 & \multicolumn{1}{c|}{15.91}                 & \multicolumn{1}{c|}{2.27}                  & \multicolumn{1}{c|}{21.74}                 & 4.35                  & \multicolumn{1}{c|}{52.86}                 & \multicolumn{1}{c|}{47.14}                 & \multicolumn{1}{c|}{25.0}                  & \multicolumn{1}{c|}{22.73}                 & \multicolumn{1}{c|}{42.86}                 & 35.71                 \\ \hline
\textbf{\begin{tabular}[c]{@{}l@{}}Claude Opus 4.1\\ (+ CodeRAG)\end{tabular}}   & \multicolumn{1}{c|}{64.52}                 & \multicolumn{1}{c|}{61.29}                 & \multicolumn{1}{c|}{50.0}                  & \multicolumn{1}{c|}{45.83}                 & \multicolumn{1}{c|}{35.0}                  & 30.0                  & \multicolumn{1}{c|}{43.59}                 & \multicolumn{1}{c|}{38.46}                 & \multicolumn{1}{c|}{56.82}                 & \multicolumn{1}{c|}{40.91}                 & \multicolumn{1}{c|}{39.13}                 & 21.74                 & \multicolumn{1}{c|}{51.43}                 & \multicolumn{1}{c|}{45.71}                 & \multicolumn{1}{c|}{25.0}                  & \multicolumn{1}{c|}{22.73}                 & \multicolumn{1}{c|}{42.86}                 & 35.71                 \\ \hline
\textbf{\begin{tabular}[c]{@{}l@{}}Claude Sonnet 4.5\\ (+ CodeRAG)\end{tabular}} & \multicolumn{1}{c|}{45.16}                 & \multicolumn{1}{c|}{41.94}                 & \multicolumn{1}{c|}{37.5}                  & \multicolumn{1}{c|}{20.83}                 & \multicolumn{1}{c|}{5.0}                   & 0.0                   & \multicolumn{1}{c|}{15.38}                 & \multicolumn{1}{c|}{12.82}                 & \multicolumn{1}{c|}{20.45}                 & \multicolumn{1}{c|}{15.91}                 & \multicolumn{1}{c|}{34.78}                 & 21.74                 & \multicolumn{1}{c|}{47.14}                 & \multicolumn{1}{c|}{45.71}                 & \multicolumn{1}{c|}{22.73}                 & \multicolumn{1}{c|}{18.18}                 & \multicolumn{1}{c|}{35.71}                 & 35.71                 \\ \hline
\textbf{Qwen Code}                                                               & \multicolumn{1}{c|}{58.06}                 & \multicolumn{1}{c|}{51.61}                 & \multicolumn{1}{c|}{62.5}                  & \multicolumn{1}{c|}{45.83}                 & \multicolumn{1}{c|}{60.0}                  & 30.0                  & \multicolumn{1}{c|}{39.74}                 & \multicolumn{1}{c|}{28.21}                 & \multicolumn{1}{c|}{50.0}                  & \multicolumn{1}{c|}{38.64}                 & \multicolumn{1}{c|}{43.48}                 & 21.74                 & \multicolumn{1}{c|}{5.71}                  & \multicolumn{1}{c|}{5.71}                  & \multicolumn{1}{c|}{4.55}                  & \multicolumn{1}{c|}{2.27}                  & \multicolumn{1}{c|}{0.0}                   & 0.0                   \\ \hline
\textbf{Claude Code}                                                             & \multicolumn{1}{c|}{87.1}                  & \multicolumn{1}{c|}{67.74}                 & \multicolumn{1}{c|}{75.0}                  & \multicolumn{1}{c|}{50.0}                  & \multicolumn{1}{c|}{60.0}                  & 50.0                  & \multicolumn{1}{c|}{41.03}                 & \multicolumn{1}{c|}{30.77}                 & \multicolumn{1}{c|}{56.82}                 & \multicolumn{1}{c|}{40.91}                 & \multicolumn{1}{c|}{43.48}                 & 30.43                 & \multicolumn{1}{c|}{57.14}                 & \multicolumn{1}{c|}{50.0}                  & \multicolumn{1}{c|}{38.64}                 & \multicolumn{1}{c|}{27.27}                 & \multicolumn{1}{c|}{35.71}                 & 28.57                 \\ \hline
\textbf{TRAE}                                                                    & \multicolumn{1}{c|}{87.1}                  & \multicolumn{1}{c|}{58.06}                 & \multicolumn{1}{c|}{79.17}                 & \multicolumn{1}{c|}{41.67}                 & \multicolumn{1}{c|}{65.0}                  & 55.0                  & \multicolumn{1}{c|}{46.84}                 & \multicolumn{1}{c|}{22.78}                 & \multicolumn{1}{c|}{54.55}                 & \multicolumn{1}{c|}{18.18}                 & \multicolumn{1}{c|}{52.17}                 & 8.7                   & \multicolumn{1}{c|}{65.71}                 & \multicolumn{1}{c|}{44.29}                 & \multicolumn{1}{c|}{54.55}                 & \multicolumn{1}{c|}{36.36}                 & \multicolumn{1}{c|}{28.57}                 & 28.57                 \\ \hline
\textbf{\oursys}                                                  & \multicolumn{1}{c|}{\textbf{96.77}}        & \multicolumn{1}{c|}{\textbf{83.87}}        & \multicolumn{1}{c|}{\textbf{79.17}}        & \multicolumn{1}{c|}{\textbf{58.33}}        & \multicolumn{1}{c|}{\textbf{90.0}}         & \textbf{85.0}         & \multicolumn{1}{c|}{\textbf{71.79}}        & \multicolumn{1}{c|}{\textbf{65.38}}        & \multicolumn{1}{c|}{\textbf{90.91}}        & \multicolumn{1}{c|}{\textbf{56.82}}        & \multicolumn{1}{c|}{\textbf{78.26}}        & \textbf{39.13}        & \multicolumn{1}{c|}{\textbf{75.71}}        & \multicolumn{1}{c|}{\textbf{70.0}}         & \multicolumn{1}{c|}{\textbf{59.09}}        & \multicolumn{1}{c|}{\textbf{47.73}}        & \multicolumn{1}{c|}{\textbf{64.29}}        & \textbf{57.14}        \\ \hline
\end{tabular}
}
\end{table*}

\hi{Performance by Synthesis Difficulty.}
We classify functions into groups of three synthesis difficulty (i.e., EASY, MEDIUM, and HARD), and assess the synthesis accuracy within each group.
We determine the synthesis difficulty of each function based on two code characteristics: (1) the number of involved function units and code tokens, and (2) the number of utilized reference function units.
The boundaries of the three groups are automatically determined using the k-means algorithm based on the derived code statistics.
As shown in Table~\ref{tab:category}, we have two observations.

\emph{First, \oursys~consistently outperforms other methods across three difficulty levels.}
Specifically, it achieves synthesis accuracies $Acc_{EXE}$ and $Acc_{RES}$ of 78.44\% and 62.60\%, which are 133.06\% and 150.80\% higher than other methods on average.
Notably, for HARD functions, \oursys~reaches an accuracy of 68.97\%, outperforming other methods by 197.10\%.
\emph{Second, the performance of \llm-based and RAG-based methods degrades severely as synthesis difficulty increases from EASY to HARD.}
Specifically, these methods achieve $Acc_{EXE}$ and $Acc_{RES}$ of 35.30\% and 32.16\% for EASY functions, while their accuracy decreases to 22.02\% and 18.90\% for HARD functions on average.
The superior performance of \oursys~is attributed to its adaptive orchestration, which efficiently navigates the database repository and manages function synthesis across varying complexities.
For instance, \oursys~implements EASY math functions in DuckDB by wrapping standard math library, such as \emph{std::cbrt} for the \emph{CbRtOperator} struct and \emph{std::pow} for the \emph{PowOperator} struct.
For HARD functions, such as \emph{bit\_count()} in DuckDB, \oursys~uses six overloaded functions, each specialized for a distinct input type (TINYINT, SMALLINT, INTEGER, BIGINT, HUGEINT, and BIT), employing appropriate operators to efficiently count the number of set bits in the binary representation.
In contrast, \llm-based and RAG-based methods primarily rely on the provided inputs and their pretrained internal knowledge, which limits their ability to accurately integrate multiple function units and manage numerous references, making them prone to errors when synthesizing HARD functions with extensive code tokens and complex references.

\begin{table}[!t]
\caption{Code Synthesis Accuracy (\%) over Function Category.}
\label{tab:category}
\resizebox{\linewidth}{!}{
\begin{tabular}{|l|cc|cc|cc|cc|}
\hline
\multicolumn{1}{|c|}{\multirow{2}{*}{\textbf{Method}}}                           & \multicolumn{2}{c|}{\textbf{Math Func.}}                           & \multicolumn{2}{c|}{\textbf{Date Func.}}                           & \multicolumn{2}{c|}{\textbf{String Func.}}                         & \multicolumn{2}{c|}{\textbf{JSON Func.}}                           \\ \cline{2-9} 
\multicolumn{1}{|c|}{}                                                           & \multicolumn{1}{c|}{\textbf{$\bm{Acc_{EXE}}$}} & \textbf{$\bm{Acc_{RES}}$} & \multicolumn{1}{c|}{\textbf{$\bm{Acc_{EXE}}$}} & \textbf{$\bm{Acc_{RES}}$} & \multicolumn{1}{c|}{\textbf{$\bm{Acc_{EXE}}$}} & \textbf{$\bm{Acc_{RES}}$} & \multicolumn{1}{c|}{\textbf{$\bm{Acc_{EXE}}$}} & \textbf{$\bm{Acc_{RES}}$} \\ \hline
\textbf{Qwen3-Coder-Plus}                                                        & \multicolumn{1}{c|}{6.67}                  & 6.67                  & \multicolumn{1}{c|}{23.53}                 & 23.53                 & \multicolumn{1}{c|}{51.35}                 & 8.11                  & \multicolumn{1}{c|}{0.0}                   & 0.0                   \\ \hline
\textbf{GPT-5}                                                                   & \multicolumn{1}{c|}{6.67}                  & 6.67                  & \multicolumn{1}{c|}{17.65}                 & 11.76                 & \multicolumn{1}{c|}{10.81}                 & 2.7                   & \multicolumn{1}{c|}{0.0}                   & 0.0                   \\ \hline
\textbf{Claude Opus 4.1}                                                         & \multicolumn{1}{c|}{6.67}                  & 6.67                  & \multicolumn{1}{c|}{29.41}                 & 23.53                 & \multicolumn{1}{c|}{24.32}                 & 16.22                 & \multicolumn{1}{c|}{4.55}                  & 4.55                  \\ \hline
\textbf{Claude Sonnet 4.5}                                                       & \multicolumn{1}{c|}{16.67}                 & 16.67                 & \multicolumn{1}{c|}{29.41}                 & 23.53                 & \multicolumn{1}{c|}{24.32}                 & 18.92                 & \multicolumn{1}{c|}{9.09}                  & 9.09                  \\ \hline
\textbf{\begin{tabular}[c]{@{}l@{}}Qwen3-Coder-Plus\\ (+ CodeRAG)\end{tabular}}  & \multicolumn{1}{c|}{40.0}                  & 6.67                  & \multicolumn{1}{c|}{35.29}                 & 29.41                 & \multicolumn{1}{c|}{27.03}                 & 16.22                 & \multicolumn{1}{c|}{40.91}                 & 40.91                 \\ \hline
\textbf{\begin{tabular}[c]{@{}l@{}}GPT-5\\ (+ CodeRAG)\end{tabular}}             & \multicolumn{1}{c|}{16.67}                 & 10.0                  & \multicolumn{1}{c|}{41.18}                 & 29.41                 & \multicolumn{1}{c|}{18.92}                 & 8.11                  & \multicolumn{1}{c|}{54.55}                 & 36.36                 \\ \hline
\textbf{\begin{tabular}[c]{@{}l@{}}Claude Opus 4.1\\ (+ CodeRAG)\end{tabular}}   & \multicolumn{1}{c|}{83.33}                 & 76.67                 & \multicolumn{1}{c|}{41.18}                 & 17.65                 & \multicolumn{1}{c|}{54.05}                 & 35.14                 & \multicolumn{1}{c|}{72.73}                 & 63.64                 \\ \hline
\textbf{\begin{tabular}[c]{@{}l@{}}Claude Sonnet 4.5\\ (+ CodeRAG)\end{tabular}} & \multicolumn{1}{c|}{23.33}                 & 23.33                 & \multicolumn{1}{c|}{41.18}                 & 29.41                 & \multicolumn{1}{c|}{35.14}                 & 21.62                 & \multicolumn{1}{c|}{9.09}                  & 9.09                  \\ \hline
\textbf{Qwen Code}                                                               & \multicolumn{1}{c|}{56.67}                 & 43.33                 & \multicolumn{1}{c|}{58.82}                 & 41.18                 & \multicolumn{1}{c|}{45.95}                 & 32.43                 & \multicolumn{1}{c|}{86.36}                 & 54.55                 \\ \hline
\textbf{Claude Code}                                                             & \multicolumn{1}{c|}{56.67}                 & 46.67                 & \multicolumn{1}{c|}{58.82}                 & 23.53                 & \multicolumn{1}{c|}{51.35}                 & 45.95                 & \multicolumn{1}{c|}{90.91}                 & \textbf{68.18}        \\ \hline
\textbf{TRAE}                                                                    & \multicolumn{1}{c|}{53.33}                 & 23.33                 & \multicolumn{1}{c|}{64.71}                 & 29.41                 & \multicolumn{1}{c|}{63.16}                 & 18.42                 & \multicolumn{1}{c|}{\textbf{95.45}}        & 40.91                 \\ \hline
\textbf{\oursys}                                                  & \multicolumn{1}{c|}{\textbf{96.67}}        & \textbf{80.0}         & \multicolumn{1}{c|}{\textbf{94.12}}        & \textbf{52.94}        & \multicolumn{1}{c|}{\textbf{89.19}}        & \textbf{62.16}        & \multicolumn{1}{c|}{\textbf{95.45}}        & 63.64                 \\ \hline
\end{tabular}
}
\vspace{-.45cm}
\end{table}

\begin{sloppypar}
\hi{Performance by Function Category.} 
We classify functions into category groups and assess accuracy within each group.
Table~\ref{tab:category} reports the results for four primary categories in PostgreSQL.
We observe that \oursys~consistently achieves higher and more stable synthesis accuracy across different categories.
Specifically, $Acc_{EXE}$ ranges from 89.19\% to 96.67\% across the four categories, with an overall accuracy improvement of 151.11\% over other methods.
This advantage arises from \oursys~explicitly leveraging function category information.
Unlike other methods that apply a uniform framework from scratch, \oursys~identifies distinctive function units through pairwise code comparison and retrieves relevant reference units within the same category.
For example, it provides \llm~with string-related reference units, including \emph{pg\_database\_encoding\_max\_length}, \emph{PG\_GETARG\_TEXT\_PP}, and \emph{PG\_RETURN\_TEXT\_P}, to implement string functions such as \emph{text\_substr()} for SQL-level function \emph{substr()} and \emph{text\_reverse()} for SQL-level function \emph{reverse()} in \emph{src/backend/utils/adt/varlena.c}.
In contrast, \llm-based methods rely exclusively on pretrained knowledge, preventing them from fully exploiting category-specific patterns. Similarly, agent-based methods are inefficient because they repeatedly search the repository from scratch (e.g., scanning files based on keywords) without reusing information from functions in the same category, reducing their overall effectiveness.
\end{sloppypar}

\subsection{Ablation Study}
\label{subsec:ablation}

We assess the synthesis accuracy of five variants over the three modules in \oursys.
\emph{(1) Function Code Characterization:} We remove the collected information in the \llms context, including the identified distinct function units, and the potential reference function units;
\emph{(2) Function Code Synthesis Operation:} We alternatively remove two operations (i.e., Pseudo-based Plan Generation, and Progressive Code Validation);
\emph{(3) Adaptive Tool Orchestration:} We replace adaptive tool orchestration with a fixed multi-\llm~collaboration without repository exploration.
Table~\ref{tab:ablation} reports the results on three databases and we have three observations below.

\begin{table}[!t]
\caption{Code Synthesis Accuracy (\%) of \oursys~Variants.}
\label{tab:ablation}
\resizebox{\linewidth}{!}{
\begin{tabular}{|cc|cc|cc|cc|}
\hline
\multicolumn{2}{|c|}{\textbf{Component}}                                                                                                                         & \multicolumn{2}{c|}{\textbf{SQLite}}                                            & \multicolumn{2}{c|}{\textbf{PostgreSQL}}                                        & \multicolumn{2}{c|}{\textbf{DuckDB}}                                            \\ \hline
\multicolumn{2}{|c|}{\textbf{\begin{tabular}[c]{@{}c@{}}Function Code Characterization\end{tabular}}}                                                          & \textbf{$\bm{Acc_{EXE}}$}                  & \textbf{$\bm{Acc_{RES}}$}                  & \textbf{$\bm{Acc_{EXE}}$}                  & \textbf{$\bm{Acc_{RES}}$}                  & \textbf{$\bm{Acc_{EXE}}$}                  & \textbf{$\bm{Acc_{RES}}$}                  \\ \hline
\multicolumn{2}{|c|}{\textbf{$\times$}}                                                                                                                          & 68.0                                   & 54.67                                  & 31.25                                  & 16.07                                  & 44.90                                  & 28.57                                  \\ \hline
\multicolumn{2}{|c|}{\textbf{\begin{tabular}[c]{@{}c@{}}Function Code\\ Synthesis Operation\end{tabular}}}                                                       & \multirow{2}{*}{\textbf{$\bm{Acc_{EXE}}$}} & \multirow{2}{*}{\textbf{$\bm{Acc_{RES}}$}} & \multirow{2}{*}{\textbf{$\bm{Acc_{EXE}}$}} & \multirow{2}{*}{\textbf{$\bm{Acc_{RES}}$}} & \multirow{2}{*}{\textbf{$\bm{Acc_{EXE}}$}} & \multirow{2}{*}{\textbf{$\bm{Acc_{RES}}$}} \\ \cline{1-2}
\textbf{\begin{tabular}[c]{@{}c@{}}Pseudo-based\\ Plan Generation\end{tabular}} & \textbf{\begin{tabular}[c]{@{}c@{}}Three-Stage\\ Code Validation\end{tabular}} &                                        &                                        &                                        &                                        &                                        &                                        \\ \hline
\textbf{$\times$}                                                               & \textbf{\checkmark}                                             & 74.67                                  & 57.33                                  & 37.04                                  & 20.37                                  & 48.98                                  & 30.61                                  \\
\textbf{\checkmark}                                              & \textbf{$\times$}                                                              & 38.67                                  & 32.0                                   & 6.9                                    & 5.52                                   & 34.69                                  & 28.57                                  \\
\textbf{$\times$}                                                               & \textbf{$\times$}                                                              & 18.67                                  & 17.33                                  & 9.66                                   & 7.59                                   & 26.53                                  & 14.29                                  \\ \hline
\multicolumn{2}{|c|}{\textbf{\begin{tabular}[c]{@{}c@{}}Adaptive Tool Orchestration\end{tabular}}}                                                             & \textbf{$\bm{Acc_{EXE}}$}                  & \textbf{$\bm{Acc_{RES}}$}                  & \textbf{$\bm{Acc_{EXE}}$}                  & \textbf{$\bm{Acc_{RES}}$}                  & \textbf{$\bm{Acc_{EXE}}$}                  & \textbf{$\bm{Acc_{RES}}$}                  \\ \hline
\multicolumn{2}{|c|}{\textbf{$\times$}}                                                                                                                          & 65.33                                  & 49.33                                  & 29.76                                  & 21.19                                  & 51.02                                  & 30.61                                  \\ \hline
\multicolumn{2}{|c|}{\textbf{\oursys (Ours)}}                                                                                                     & \textbf{81.33}                         & \textbf{69.33}                         & \textbf{78.62}                         & \textbf{58.62}                         & \textbf{83.67}                         & \textbf{67.35}                         \\ \hline
\end{tabular}
}
\end{table}

\emph{First, all proposed components are essential for accurate synthesis, with their removal leading to different degrees of accuracy drops.}
\oursys~obtains the improvement of $Acc_{EXE}$ and $Acc_{RES}$ by 42.14\% and 37.50\% with these components on average.
For example, the function code characterization module successfully pinpoints the three required function units (i.e., \emph{sumStep}, \emph{totalFinalize}, and \emph{sumInverse}) for the \emph{total()} window aggregate function in SQLite.
In the synthesis operation module, pseudo-based plan generation highlights potential reference units to guide function synthesis, preventing errors such as ``unknown type name MultiRangeType'' in PostgreSQL's \emph{lower()} function.
The adaptive tool orchestration enables progressive correction of errors through iterative code synthesis operations, resolving issues such as \emph{``unknown jsonb value type''} detected during the validation of \emph{``SELECT json\_pretty()''}, and ultimately producing the ground-truth implementation.

\emph{Second, progressive code validation is crucial for reliable synthesis, and removing it leads to significant performance decline.} 
For complex databases such as PostgreSQL, removing the three-stage validation causes a significant accuracy drop, with $Acc_{EXE}$ falling from 78.62\% to 6.9\%. 
This drop reflects PostgreSQL's complexity (rather than \oursys~being specifically tuned for PostgreSQL), with more cross-file dependencies and more multi-branch execution paths than SQLite, where three-stage validation is vital for resolving diverse internal dependencies and verifying complex execution logic.
Without feedback from the three-stage validation, \oursys~must rely on the \llm's pretrained knowledge (increasing the risk of producing incorrect or hallucinated macros) or coarse shell commands (ineffective for identifying deeper semantic failures). 
For instance, when synthesizing \emph{date\_trunc()}, (1) syntax and compliance-level validations identify the misuse of the \emph{TIMESTAMP\_GET\_FIELD()} macro and missing \emph{pg\_proc} registration entries, while (2) semantic-level validation detects logic flaws in handling negative intervals. 
These feedbacks enable the \llm~to iteratively refine its initial code implementations.

\emph{Third, components such as the pseudo-based plan generation and progressive validation work together to handle complex logic, and using them together improves accuracy.}
For example, when planning is disabled, PostgreSQL's $Acc_{RES}$ increases slightly (5.52\% $\rightarrow$ 7.59\%), but this does not imply that planning is harmful.
Instead, it reflects that coding plans and validations are interdependent for complex systems such as PostgreSQL.
The coding plan is designed to be comprehensive, listing potential code elements (e.g., logic branches) to ensure \llm does not miss essential logic.
For example, the plan for \emph{date\_trunc()} includes logic for handling \emph{tsrange} inputs (e.g., extracting range bounds), which are invalid for standard \emph{timestamp} inputs. Without validation to remove these errors, \llm would try to apply this range logic to standard inputs, causing type errors that the validation would catch.
When the plan is disabled, \llm avoids these errors via its pre-trained knowledge but still misses logic branches (e.g., for \emph{interval} types).
The absolute difference is small (2.07\%) compared to \oursys's improvement (37.50\%), showing that turning off these components fails to handle complex systems.

\subsection{New Native Function Synthesis}
\label{subsec:newfunc}

\begin{table*}[!t]
\caption{Extending SQLite with New Native Functions Generated by \oursys (See code in our \href{https://github.com/weAIDB/DBCooker}{\gray{[\underline{artifact}]}}).}
\label{tab:newfunc}
\resizebox{\linewidth}{!}{
\begin{tabular}{|c|c|c|c|c|c|c|c|}
\hline
\rowcolor[HTML]{000000} 
{\color[HTML]{FFFFFF} \textbf{Name}} & {\color[HTML]{FFFFFF} \textbf{Category}} & {\color[HTML]{FFFFFF} \textbf{Source}} & {\color[HTML]{FFFFFF} \textbf{Declaration}}                                                                                                                         & {\color[HTML]{FFFFFF} \textbf{Qwen Code}} & {\color[HTML]{FFFFFF} \textbf{Claude Code}} & {\color[HTML]{FFFFFF} \textbf{TRAE}} & {\color[HTML]{FFFFFF} \textbf{\oursys}} \\ \hline
\textbf{covar\_pop}                  & Aggregate Function                       & PostgreSQL                             & \begin{tabular}[c]{@{}c@{}}WAGGREGATE(covar\_pop, 2,0,0, covarPopStep, \\ covarPopFinalize, covarPopFinalize, covarPopInverse, 0)\end{tabular}                      & $\times$                                & \checkmark                                  & \checkmark                           & \checkmark                                             \\ \hline
\textbf{bool\_and}                   & Aggregate Function                       & PostgreSQL                             & \begin{tabular}[c]{@{}c@{}}WAGGREGATE(bool\_and, 1, 0, 0, boolAndStep, \\ boolAndFinalize, boolAndFinalize, boolAndInverse, \\ SQLITE\_FUNC\_ANYORDER)\end{tabular} & $\times$                                & \checkmark                                  & \checkmark                           & \checkmark                                             \\ \hline
\textbf{bool\_or}                    & Aggregate Function                       & PostgreSQL                             & \begin{tabular}[c]{@{}c@{}}WAGGREGATE(bool\_or, 1, 0, 0, boolOrStep, \\ boolOrFinalize, boolOrValue, boolOrInverse, \\ SQLITE\_FUNC\_ANYORDER)\end{tabular}         & \checkmark                                & $\times$                                  & \checkmark                           & \checkmark                                             \\ \hline
\textbf{century}                     & Date Function                            & DuckDB                                 & FUNCTION(century,  1, 0, 0, centuryFunc)                                                                                                                            & $\times$                                & $\times$                                  & $\times$                           & \checkmark                                             \\ \hline
\textbf{monthname}                   & Date Function                            & DuckDB                                 & FUNCTION(monthname, 1, 0, 0, monthnameFunc)                                                                                                                         & $\times$                                & $\times$                                  & $\times$                           & \checkmark                                             \\ \hline
\textbf{yearweek}                    & Date Function                            & DuckDB                                 & PURE\_DATE(yearweek, 1, 0, 0, yearweekFunc)                                                                                                                         & $\times$                                & \checkmark                                  & \checkmark                           & \checkmark                                             \\ \hline
\textbf{last\_day}                   & Date Function                            & DuckDB                                 & FUNCTION(last\_day, 1, 0, 0, last\_dayFunc)                                                                                                                         & \checkmark                                & $\times$                                  & \checkmark                           & \checkmark                                             \\ \hline
\textbf{lcm}                         & Numeric Function                         & DuckDB                                 & FUNCTION(lcm, 2, 0, 0, lcmFunc)                                                                                                                                     & \checkmark                                & \checkmark                                  & $\times$                           & \checkmark                                             \\ \hline
\textbf{even}                        & Numeric Function                         & DuckDB                                 & FUNCTION(even, 1, 0, 0, evenFunc)                                                                                                                                   & $\times$                                & $\times$                                  & $\times$                           & \checkmark                                             \\ \hline
\textbf{gamma}                       & Numeric Function                         & DuckDB                                 & FUNCTION(gamma, 1, 0, 0, gammaFunc)                                                                                                                                 & $\times$                                & $\times$                                  & $\times$                           & \checkmark                                             \\ \hline
\textbf{lgamma}                      & Numeric Function                         & DuckDB                                 & FUNCTION(lgamma, 1, 0, 0, lgammaFunc)                                                                                                                               & $\times$                                & $\times$                                  & $\times$                           & \checkmark                                             \\ \hline
\textbf{nextafter}                   & Numeric Function                         & DuckDB                                 & FUNCTION(nextafter, 2, 0, 0, nextafterFunc)                                                                                                                         & $\times$                                & $\times$                                  & $\times$                           & \checkmark                                             \\ \hline
\textbf{left}                        & String Function                          & PostgreSQL                             & FUNCTION(left, 2, 0, 0, leftFunc)                                                                                                                                   & $\times$                                & \checkmark                                  & \checkmark                           & \checkmark                                             \\ \hline
\textbf{regexp\_split\_to\_array}    & String Function                          & PostgreSQL                             & \begin{tabular}[c]{@{}c@{}}FUNCTION(regexp\_split\_to\_array, 2, 0, 0, \\ regexpSplitToArrayFunc)\end{tabular}                                                      & $\times$                                & $\times$                                  & $\times$                           & \checkmark                                             \\ \hline
\textbf{repeat}                      & String Function                          & PostgreSQL                             & FUNCTION(repeat, 2, 0, 0, repeatFunc)                                                                                                                               & \checkmark                                & \checkmark                                  & $\times$                           & \checkmark                                             \\ \hline
\textbf{to\_hex}                     & String Function                          & PostgreSQL                             & FUNCTION(to\_hex, 1, 0, 0, toHexFunc)                                                                                                                               & $\times$                                & $\times$                                  & \checkmark                           & \checkmark                                             \\ \hline
\textbf{translate}                   & String Function                          & PostgreSQL                             & FUNCTION(translate, 3, 0, 0, translateFunc)                                                                                                                         & $\times$                                & $\times$                                  & $\times$                           & \checkmark                                             \\ \hline
\end{tabular}
}
\end{table*}

\begin{sloppypar}
To evaluate \oursys~on new function synthesis, we extend SQLite's capabilities by introducing new functions from other databases.
Specifically, we collect function declarations (e.g., name, category, and SQL examples from PostgreSQL and DuckDB) to simulate new functionality requests in SQLite.
The function codes are not used as references, since native functions differ greatly across databases (see Section~\ref{sec:offline}), making direct code transfer impractical.
Table~\ref{tab:newfunc} lists the newly synthesized SQLite functions, where \oursys~presents superior performance for three reasons.
\end{sloppypar}


\emph{First, \oursys~accurately implements all required function units for a new function and declares it correctly.}
It realizes the core processing logic and enables SQL interface by declaring the function.
For example, it implements three required function units for the SQLite \emph{covar\_pop()} aggregate function, including \emph{covarPopStep} to update sums and counts, \emph{covarPopFinalize} to compute the final covariance, and \emph{covarPopInverse} to perform inverse calculations.
The function is then declared with \emph{WAGGREGATE(covar\_pop, ..., covarPopStep, covarPopFinalize, ..., covarPopInverse, 0)} in \emph{src/func.c}, making it callable via \emph{SELECT covar\_pop()}.
In contrast, Qwen Code declares but fails to implement \emph{covarPopInverse}, introducing compliance errors \emph{``undeclared `xInverse' in definition of macro `WAGGREGATE' ''}.


\emph{Second, \oursys~effectively leverages reference function units within the database repository to support accurate function synthesis.}
It correctly incorporates both standard and functional reference units in each synthesized function. 
For example, in the \emph{bool\_or} function, \oursys~uses \emph{sqlite3\_aggregate\_context} for context management and \emph{sqlite3\_value\_type(argv[0]) != SQLITE\_NULL} for input validation within \emph{boolOrStep}. 
In contrast, Claude Code introduces a redefinition error for `struct SumCtx' during synthesis. 
Furthermore, \oursys~successfully implements the \emph{regexp\_split\_to\_array} function by utilizing reference units such as \emph{sqlite3\_value\_text} and \emph{sqlite3\_str\_new}, whereas the other three methods encounter undefined references to `sqlite3re\_compile' and `sqlite3re\_match' in the external \emph{sqlite3re.c}, resulting in compliance errors.

\emph{Third, \oursys~adaptively manages diverse functions and progressively refines code to achieve the intended functionality.}
It determines appropriate declarations and generates function codes for diverse functions through adaptive tool orchestration as shown in Table~\ref{tab:newfunc}.
For instance, it differentiates aggregate functions from other scalar functions using distinct macros (e.g., \emph{WAGGREGATE} vs. \emph{FUNCTION}) and assigns specific function names and arguments according to the category (e.g., \emph{centuryFunc} with a single argument for the date function \emph{century()}).
In contrast, other methods rely on a uniform synthesis process, exploring the database repository from scratch, which reduces effectiveness.
For example, Claude Code fails to generate the \emph{century} function, and TRAE fails to generate \emph{translate} due to exhaustive and irrelevant repository searches.

\section{Related Work}
\label{sec:related}

\hi{General Code Generation.} Recent advances in code generation are broadly categorized into three categories:
(1) Prompt-based methods (e.g., Codex~\cite{chen2021evaluating}, GitHub Copilot~\cite{copilot2023empirical}) treat code generation as conditional text generation from natural language or code prompts;
(2) Agent-based systems (e.g., Claude Code~\cite{claudecode}, SWE-Dev~\cite{qian2023sweagent}, Qwen Code~\cite{qwencode}) enhance LLMs with planning and tool usage for multi-step reasoning and code debugging;
(3) Training-based models (e.g., Code Llama~\cite{roziere2023codellama}, StarCoder~\cite{li2023starcoder}, WizardCoder~\cite{luo2023wizardcoder}) focus on architectural improvements via pretraining on large code corpora or reinforcement learning from execution traces. Despite these advances, they are still limited in complex database-native function synthesis (see results in Section~\ref{sec:exp}), which requires precise file searches, function-level references, and adaptive synthesis across varying input types and database constraints.


\hi{UDF Optimization.}
Existing work on User-Defined Function (UDF) optimization focuses on transforming high-level UDF logic into efficient internal representations.
Froid~\cite{Froid} converts imperative UDFs into relational algebraic expressions, embedding them into SQL for cost-based optimization and parallel execution.
Tuplex~\cite{Tuplex} accelerates and compiles Python UDFs into native code, with a dual-mode strategy that aggressively optimizes the common case and handles exceptions through a fallback path.
Unlike \oursys, these methods focus on logic rewriting or runtime compilation rather than kernel-level code implementation.

\hi{Runtime Code Generation.}
Existing studies accelerate execution through temporary machine code. 
HyPer~\cite{HyPer} compiles queries into LLVM-based machine code using a data-centric push-based model that retains data in CPU registers for optimized locality and branch prediction. 
Weld~\cite{Weld} employs a common intermediate representation for cross-library workflows, minimizes data movement, and generates efficient parallel code. 
These approaches differ from \oursys, focusing on temporary code for immediate execution rather than reusable, persistent native functions.

\hi{Database Migration.}
Recent research uses \llms to bridge gaps across SQL dialects.
CrackSQL~\cite{cracksql, cracksqldemo} automates dialect translation via a local-to-global strategy, iteratively refining translations by fixing local execution failures.
PARROT~\cite{parrot} proposes a benchmark for cross-system SQL translation, specifically testing system-specific syntax. 
These works highlight the complexity and manual effort required to reimplement native functions across database systems.

\section{Conclusion and Future Work}
\label{sec:conclusion}

In this paper, we present \oursys, the first \llm-based system for automatic database native function code synthesis.
It proposes a function characterization module to capture function declarations, distinctive units, and cross-unit references, with three database-aware synthesis operations (pseudo-based plan generation, progressive code synthesis, and three-stage code validation) and adaptive tool orchestration for diverse function synthesis.
Experiments on three databases show that \oursys outperforms state-of-the-art methods and effectively supports new function synthesis. 

As frontier models advance, \oursys~can continue to strengthen their capabilities for more effective function synthesis.

\textbf{(1) Massive Codebase Fragmentation vs. Long-Context Reasoning:} 
Database codebases are massive (e.g., over one million lines of code in PostgreSQL~\cite{postgresqlfunction}) with scattered function units (e.g., separating declarations in \emph{pg\_proc.dat} from implementations in \emph{src/func.c}), making it difficult to fully include in \llms' context window.
Even if future \llms can process the entire codebase, blindly including everything incurs high inference costs, and effective synthesis still requires reliably identifying a small set of relevant, scattered units among a vast amount of irrelevant code.
\oursys's \textit{Function Characterization} explicitly identifies these units, preventing \llms from missing critical information that long-context reasoning over the entire codebase alone may overlook~\cite{LostInMiddle, LongCodeBench}.

\textbf{(2) Deterministic Correctness Requirements vs. Probabilistic Generative Synthesis:} 
Database systems are mission-critical, and their functions must satisfy strict, deterministic correctness guarantees.
Even if future \llms become more reliable, their outputs are still based on probabilistic generation and cannot inherently guarantee exact database correctness~\cite{ProbabilisticLLM, CodeT}.
\oursys's \textit{Function Code Synthesis Operations} remain essential by explicitly enforcing structural templates and correctness constraints via pseudo-plans and three-stage progressive validation.
This external enforcement improves reliability by turning probabilistic generation into verifiable and correct implementations, instead of relying on \llms alone to ensure strict database correctness.

\textbf{(3) Dynamic Codebase Adaptation vs. Static Training Bias:}
Database functions evolve across versions (e.g., changes in function signatures, required macros, or system catalog definitions)~\cite{duckdbversion}.
Even with broader and newer training data, future \llms inevitably learn a mixture of database versions and deprecated conventions, which might conflict with the exact requirements of a target database version~\cite{LibEvolutionEval, RustEvo}.
\oursys addresses this by dynamically retrieving version-specific database implementation routines and enforcing them via \textit{Adaptive Tool Orchestration}, allowing direct adaptation to database version changes without model retraining.




\begin{acks}
Xuanhe Zhou is the corresponding author.
This work was supported in part by National Key R\&D Program of China (No. 2023YFB4502400), NSF of China (No. 62502304, U25A20437, 62525202, 62232009, 62441236, 62372296, 62432007, and U25A6024), Fundamental and Interdisciplinary Disciplines Breakthrough Plan of the Ministry of Education of China (No. JYB2025XDXM103), Shenzhen Project (CJGJZD20230724093403007), CCF Populus Grove Fund, ByteDance, Tencent, Ant Group through CCF-Ant Research Fund, Shanghai Jiao Tong University AI for Engineering Initiative, Shanghai Artificial Intelligence Laboratory, Alibaba Group through Alibaba Innovation Research Program, Tencent Rhino Bird Key Research Project, China Railway Science Research Institute Group Co., Ltd, Zhongguancun Lab, Huawei, and Beijing National Research Center for Information Science and Technology (BNRist).
\end{acks}

\clearpage

\bibliographystyle{ACM-Reference-Format}
\bibliography{sample-base}

\end{document}